\documentclass[ifpdf]{ar-1col}
\usepackage{natbib}
\pdfoutput=1
\usepackage{ifpdf}
\usepackage[mathscr]{euscript}
\usepackage{mathtools}

\jname{Xxxx. Xxx. Xxx. Xxx.}
\jvol{57}
\jyear{2019}
\doi{10.1146/((please add article doi))}

\newcommand{\simkl}{\stackrel{{<}}{{\scriptstyle\sim}}}
\newcommand{\simgr}{\stackrel{{>}}{{\scriptstyle\sim}}}

\newcommand{\dxdy}[2]{{\frac{\partial{#1}}{\partial{#2}}}}

\def\xir{\xi_r}
\def\xih{\xi_{\rm h}}
\def\xirnl{\xi_{r,nl}}
\def\xihnl{\xi_{{\rm h},nl}}
\def\dd{{\rm d}}

\begin{document}

\markboth{Aerts, Mathis, Rogers}{Angular Momentum Transport in Stellar Interiors}

\title{Angular Momentum Transport in Stellar Interiors}

\author{Conny Aerts,$^{1,2}$ St\'ephane Mathis,$^{3,4}$ and Tamara M.\ Rogers$^{5,6}$
\affil{$^1$ Institute of Astronomy, KU\,Leuven, Celestijnenlaan 200D,
B-3001 Leuven, Belgium; email: Conny.Aerts@kuleuven.be}
\affil{$^2$ Department of Astrophysics, IMAPP, Radboud University
Nijmegen, P.O.\ Box 9010, 6500 GL Nijmegen, The Netherlands}
\affil{$^3$Laboratoire AIM Paris-Saclay, CEA/DRF - CNRS - Université Paris Diderot,
  IRFU/DAp Centre de Saclay, F-91191 Gif-sur-Yvette, France; Stephane.Mathis@cea.fr}
\affil{$^4$LESIA, Observatoire de Paris, PSL Research University, CNRS,
  Sorbonne Universit\'es, UPMC Universit\'e Paris 06, Universit\'e Paris
  Diderot, Sorbonne Paris Cit\'e, 5 place Jules Janssen, F-92195 Meudon, France}
\affil{$^5$Department of Mathematics, Statistics and Physics, Newcastle University,
  Newcastle upon Tyne, UK; Tamara.Rogers@newcastle.ac.uk}
\affil{$^6$Planetary Science Institute, Tucson, AZ 85721, USA}}

\begin{abstract}
  Stars lose a significant amount of angular momentum between birth and death,
  implying that efficient processes transporting it from the core to the surface
  are active.  Space asteroseismology delivered the interior rotation rates of
  more than a thousand low- and intermediate-mass stars, revealing that:
  1)\,single stars rotate nearly uniformly during the core hydrogen and core
  helium burning phases; 2)\,stellar cores spin up to a factor 10 faster than
  the envelope during the red giant phase; 3)\,the angular momentum of the
  helium-burning core of stars is in agreement with the angular momentum of
  white dwarfs.  Observations reveal a strong decrease of core angular momentum
  when stars have a convective core. Current theory of angular momentum
  transport fails to explain this.  We propose improving the theory
  with a data-driven approach, whereby angular momentum prescriptions derived
  from multi-dimensional (magneto)hydrodynamical simulations and theoretical
  considerations are continously tested against modern observations.  The TESS
  and PLATO space missions have the potential to derive the interior rotation of
  large samples of stars, including high-mass and metal-poor stars in binaries and
  clusters.  This will provide the powerful observational constraints needed to
  improve theory and simulations.
\end{abstract}

\maketitle

\tableofcontents

\section{SETTING THE STAGE}

Stars are radiating rotating gaseous spheres whose interior properties meet
the laws of gravity and hydrodynamics. Understanding the properties of stellar
interiors up to present-day observational precision requires 
a complex interplay between atomic and nuclear physics, (quantum-)mechanics and
thermodynamics, with intense interactions between radiation and matter.
Moreover, the dynamics in stellar interiors and the life cycles of stars happen
on length and time scales that span so many orders of magnitude that we remain
very far from full 3D stellar evolution computations, even with the most
powerful computers now or in
the foreseeable future. We must
therefore resort to 1D approximations when computing stellar evolution models.

Besides being fascinating in their own right, theoretical stellar structure and
evolution models are basic ingredients for many areas of
astrophysics. Indeed, stars are the building blocks of galaxies, clusters,
associations, binaries, and exoplanetary systems.  Therefore, these
studies, in addition to stellar evolution theory, rely heavily on stellar
models (and their accuracy). These 
models are constantly subjected to new observational constraints, as more, and
particularly more precise, data become available. Such has happened with
high-precision space photometry the past decade, with pleasant confirmation of
how well some aspects of stellar models represent reality, but with 
unanticipated surprises revealing how poor the models are on other aspects.

In this review, we focus on one pertinent aspect of stellar models that turns
out to be in need of major improvement: the angular momentum transport that
occurs inside a star during its evolution. We treat this topic from three
complementary aspects: modern observations, state-of-the art theory and multi-D
numerical simulations. Each of these aspects is detailed below, with brief
historical reminders. The historical path for these three aspects is quite
different in duration and maturity, with theory going back almost a century,
hydrodynamical 
simulations in 2D or 3D having started only when computational power allowed it
since some 30 years, and direct observations of stellar interiors for stars
other than the Sun and white dwarfs 
reaching sufficient precision only about a decade ago, when
space asteroseismology came into practice. 

\begin{marginnote}[]
\entry{Low-mass star} a star that, at the onset of core hydrogen burning, has
a radiative core or is fully convective ($M \simkl 1.3\,$M$_\odot$)
and will end its life as a white dwarf
\entry{Intermediate-mass star} a star that has a convective core at the onset
of core hydrogen burning
and that will end its life 
as a white dwarf ($1.3 \simkl M \simkl 8\,$M$_\odot$)
\entry{High-mass star} a star born with a large convective core and with sufficient
  mass ($M \simgr 8\,$M$_\odot$) to undergo core collapse at the end of its life
  to form a neutron star or a black hole
\end{marginnote}

When it comes to predicting the type of remnant at the end of stellar life -- a
white dwarf for low- or intermediate-mass stars born with a mass
$M \simkl 8\,$M$_\odot$ and a neutron star or black hole for high-mass stars
born with a mass $M \simgr 8\,$M$_\odot$ -- the theory of stellar evolution is
quite well established
\citep[e.g.,][for modern monographs]{Maeder2009,Kippenhahn2012}. 
On the other hand, the theory of stellar interiors relies on
physical concepts that are still subject to considerable uncertainties. Indeed,
many of the (intrinsically multi-dimensional) phenomena connected with rotation, magnetism, mixing of chemical
elements, and angular momentum transport inside stars cannot be deduced from
first principles. Such physical ingredients therefore include one or more free
parameters and these remained essentially uncalibrated prior to the
asteroseismology era, because observational measurements were mostly limited to
constraints coming from the stellar photosphere.

Long-standing theoretical concepts taken for granted for stellar structure
computations, such as local conservation of angular momentum, rotational mixing,
and dynamical instabilities \citep[see the extensive monographs by,
e.g.,][]{Hansen2004,Maeder2009} turn out to have unanticipated inaccuracies in
terms of the transport of chemical elements and of angular momentum induced by
them in the deep stellar interior. Theory and observations differ by two orders
of magnitude with regard to core-to-envelope rotation rates in low-mass
evolved stars, while the level of chemical mixing in young intermediate-mass
stars connected with instabilities in their radiative layers adjacent to the
core are orders of
magnitude lower than anticipated.  These limitations of the models will be
further outlined below. Many of these difficulties were only recently uncovered thanks to the probing
capacity of stellar oscillations, offering a direct view of deep stellar
interiors.

\begin{figure}[t!]
\begin{center}
\rotatebox{0}{\resizebox{4.5cm}{!}{\includegraphics{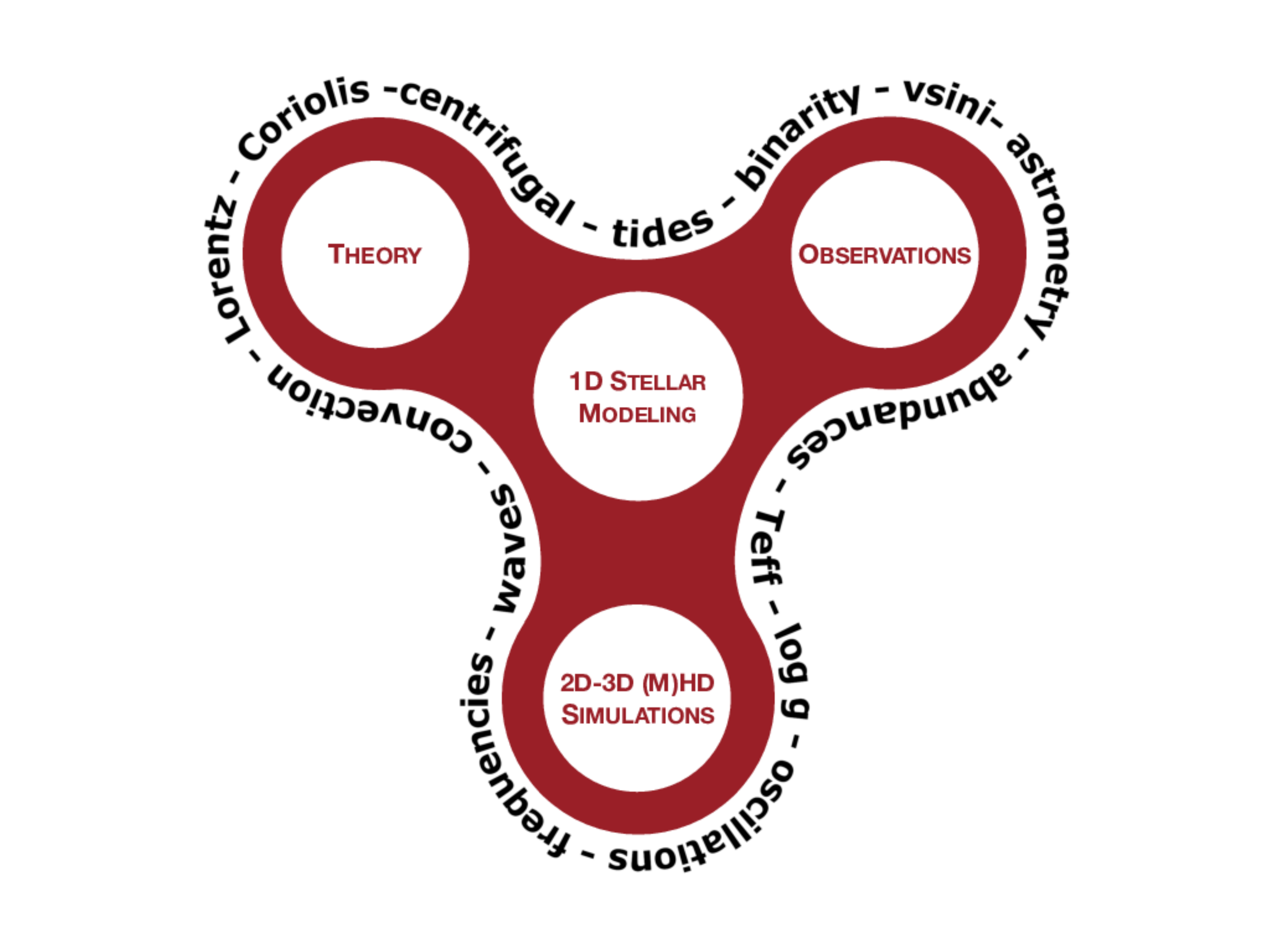}}}\vspace{-0.75cm}
\caption{\label{spinner} Synergies and complementarities between theory, various
  types of observations, and multi-D simulations. Bridging all those aspects
  will lead to an inclusive methodology for 1D stellar modeling and a
  better understanding of stellar interiors, including angular momentum
  transport.}
\end{center}
\end{figure}

This review first offers a concise reminder of ``classical'' observational
constraints from spectroscopy, interferometry, and astrometry of single, binary,
and cluster stars used to calibrate stellar interiors, after a brief discussion
of the differences between 1D stellar evolution models with and without
rotation. Such classical diagnostics at best have relative precisions of order a
few percent, but will remain an important observational method to evaluate
various types of stellar models for stars and stellar systems that do not reveal
oscillations suitable for asteroseismology.  Next, we highlight the
recent asteroseismic input from stars that offer the opportunity to calibrate
stellar models, with specific emphasis on internal rotation. Finally, we place
the new observational diagnostics into context, by providing a historical
overview and the current status of the theory of transport processes in stars,
detailing angular momentum transport in particular. We highlight the gain from a
fully integrated approach combining observations, theory, and multi-D
simultations into a global picture.  Such a comprehensive approach requires the
application and integration of a multitude of techniques and methods based on
observations and theory. An encompassing visualisation is offered in
Figure\,\ref{spinner}, where each of the keywords used to construct it will be
discussed below. It represents graphically the conclusion of our review on the
way forward towards a better understanding of angular momentum transport in
stars.

\begin{marginnote}[]
\entry{J} angular momentum of a star, defined as $M\cdot\Omega\cdot R^2$ and
expressed in the SI units kg\,m$^2$\,s$^{-1}$
\end{marginnote}

\subsection{\label{1Dmodels}1D Stellar Evolution Models}

As will be outlined in Section\,\ref{Astero-age}, we now have asteroseismic probes
of the properties of stellar interiors. Such diagnostics give direct access to
the physics at different depths inside the star, in particular the regions near
the stellar core. Stellar models relying on physical descriptions that were
taken for granted for decades, can now finally be calibrated
asteroseismically. From this, it has been found that aspects of ``well
established'' theory do not meet the asteroseismic requirements,
particulary in the case of rotation and angular momentum transport. 
In order to offer a path
towards improved theory, we first remind the reader of some basic ingredients
of stellar structure theory.

Stellar models are obtained by solving a set of differential equations based on
the laws of physics, accompanied by proper boundary conditions for the center
and surface of the star. Solving these equations requires that choices for the
microscopic properties of stellar matter are made, in particular the
equation of state of the gas, nuclear reactions, the interaction of
radiation and matter, etc.  Despite improved computational power the past
decades, stellar models necessarily remain a simplified version of reality,
because it is not possible to compute complete 3D models across stellar
evolution. Pragmatic choices in the approximations adopted to solve the stellar
structure equations are thus in order.  A major simplification is achieved when
stellar rotation and stellar magnetism are ignored. Indeed, whenever the
accompanying Coriolis, centrifugal and Lorentz forces are neglected,
spherical symmetry occurs.  Moreover, from a physical viewpoint, rotation and
magnetism introduce a multitude of flows, waves, and instabilities, with
accompanying yet uncalibrated transport processes, each having a complex
feedback on the stellar structure as time evolves
\citep{Maeder2009,Mathis2013LNP}.  It is then obvious how to simplify the
stellar model computations in first instance, namely by ignoring rotation and
magnetism. These phenomena are therefore often only considered when 1D models
fail to explain particular observational diagnostics.

\begin{marginnote}[]
\entry{Main sequence} the phase in a star's evolution when it is burning
hydrogen in its core
\end{marginnote}

How appropriate is it to ignore rotation and magnetism when computing stellar
models? Both phenomena are closely related and their importance for stellar
evolution depends on the mass of the star.  About 10\% of intermediate- and
high-mass stars have a detectable stable large-scale structured fossil magnetic field left from their formation and previous convective phases at the current
threshold of CaII H \& K absorption lines or spectropolarimetry based on Zeeman
splitting \citep[typically a few Gauss,][]{DonatiLandstreet2009,Wade2016}.  This percentage is in agreement with the 
broad distribution of the measured spectroscopic 
projected equatorial surface rotation velocity, $v\sin\,i$
\citep[e.g.,][]{ZorecRoyer2012}. The physical interpretation is that such stars
are born without or with only a thin convective envelope, preventing the
creation of a magnetic dynamo in their outer layers.  In absence of the latter,
they do not experience a strong magnetized wind and therefore do not lose much
angular momentum.  While stars born with a mass above some $\sim\!15\,$M$_\odot$
do experience a considerable radiation-driven wind \citep{Kudritzki2000} and
lose angular momentum efficiently, the majority of stars with
$1.3 \simkl M \simkl 15\,$M$_\odot$ roughly keep their original $J$ received at
birth during the main sequence.  Aside from these considerations for the stellar
envelope, high- and intermediate-mass stars may have a
magnetic dynamo in their convective core region.  The strength and properties of such
interior field cannot be measured directly, as it is shielded by the
extended overlying radiative envelope.  Indirect evidence of the existence  interior fields, with strength above $10^5$ Gauss, was deduced from
missing dipole modes in {\it Kepler} data of red giants
\citep{Fuller2015,Cantiello2016}. If these dipole modes are missing due to 
suppression by an interior magnetic field in the radiative core, 
then such field must occur in about half of the progenitor main sequence stars \citep{Stello2016}. However, \citet{Mosser2017} pointed
out that the dipole modes of these red giants are not missing, but are rather
mixed modes with depressed amplitudes due to damping, invalidating
hypotheses made by \citet{Fuller2015} and \citet{Cantiello2016}. Irrespective of this
controversy, it is a sensible approach to ignore core and surface magnetism for
the mass range $1.3 \simkl M \simkl 15\,$M$_\odot$ and to focus on rotation as
the prime cause of complexity in 1D stellar models 
\citep[e.g.,][]{Georgy2013} and in asteroseismology \citep{Buysschaert2018}. 
Additional attention must be given to
mass loss for $M\simgr 15\,$M$_\odot$.

In contrast, low-mass stars with $M \simkl 1.3\,$M$_\odot$ are left with a
considerable and deep convective envelope after their formation process,
resulting in an efficient dynamo, leading to strong cyclic variability. 
Their (surface) rotation efficiently slows
down during the main sequence due to an effective magnetized wind
connected with the envelope dynamo, as first recognized from the pioneering work
of \citet[][and references therein]{Kraft1967}. This spin-down offers a way of aging
low-mass stars observationally \citep[e.g.,][]{Barnes2003}. 

\subsubsection{1D Stellar Models without Rotation and Magnetism}

Even stellar evolution models representing non-rotating non-magnetic stars have
major challenges. A critical aspect is the type of energy transport within the
various layers inside a star. In the case where energy transport by means of the
diffusion of photons is sufficiently efficient, a radiatively stratified layer
whose thermal structure is determined by the radiative temperature gradient,
will occur.  When photons are unable to transport the energy in particular
stellar layers, macroscopic convective energy transport takes place.  The time
scale of such macroscopic motions is negligible compared to the relevant time
scales of stellar evolution. Therefore, turbulent convection in stellar models
is often assumed to be time-independent and treated by the mixing-length theory
of \citet[][MLT]{BohmVitense1958}, although several alternatives exist
\citep[e.g.,][]{Canuto1991}.  MLT assumes that the convective energy transport is taken
care of by one single type of large-size eddy with one characteric mean free path,
called the mixing length.  This parameter, here denoted as $\alpha_{\rm MLT}$,
is usually expressed in local pressure scale heights, $H_P$, and takes typical
values between 0.5 and 2.5, depending on the mass and metalicity of the star
\citep[e.g.][for recent evaluations and calibration with respect to the solar
value]{Viani2018}.  For low-mass main-sequence stars, $\alpha_{\rm MLT}$
influences the size of the outer convective envelope.  Intermediate-mass stars
with $1.3 \simkl M \simkl 2\,$M$_\odot$ have a thin convective envelope and
stars born with a higher mass mainly have convection in their fully mixed
core. The physical circumstances in these stars are very different from those in
the Sun, hence $\alpha_{\rm MLT}$ is essentially uncalibrated for such stars.

\begin{marginnote}[]
  \entry{MLT} mixing-length theory of convection
\end{marginnote}

The full and instantaneous mixing caused by convection leaves a clear mark on
the chemical history of the star.  A major unknown in the theory of stellar
interiors is the treatment of the physical conditions in the transition layers
between a convective and a radiative zone. The thermal structure, as well as the
efficiency of chemical mixing and transport of angular momentum in such
transition layers are unknown.  The size and location of these transition layers
are difficult to determine because they depend on the phenomenon of convective
``overshooting'' or ``penetration'' as discussed in Section\,\ref{penetration}.
Due to their turbulent motion and inertia, the fluid elements do not stop abruptly
when entering a radiative layer.  Their movement continues over some overshoot
depth, $\alpha_{\rm ov}$ (expressed in $H_P$; this is $d_{\rm pen}$ or
$d_{\rm over}$ for penetration or overshooting, as explained in
Section\,\ref{penetration}).  The treatment of the mixing in this overshoot zone
extending beyond the Schwarzschild boundary at radial coordinate $r_{\rm cc}$,
here denoted as $D_{\rm mix}(r\!>\!r_{\rm cc})$, is of critical importance because
it determines how much fuel takes part in the nuclear reactions. In particular,
the amount of helium/carbon at the end of the core hydrogen/helium burning,
depends on the structure, hydrodynamics, and mixing efficiency in the overshoot
zone. The observational estimation of the properties of the overshoot zone is
hence of major importance, because these determine the core mass during the
evolution of the star.  Convective envelope overshooting (aka undershooting) may
occur as well, but its effect on the star's evolution is small compared to the
one of core overshooting, although it does affect the surface chemistry.
\begin{marginnote}[]
  \entry{The mixing length} a free parameter $\alpha_{_{\rm MLT}}$ 
representing the characteristic
  length scale over which convective fluid elements travel before they dissipate
  in their environment
\end{marginnote}

\begin{marginnote}[]
\entry{Overshooting} the phenomenon of turbulent convective fluid elements 
entering a radiative zone over an unknown distance, expressed as a 
fraction $\alpha_{\rm ov}$  of the local pressure scale height
\end{marginnote}

\subsubsection{1D Stellar Models with Rotation}

A large fraction of early-F to O stars are fast rotators \citep{ZorecRoyer2012}.
Observations of such stars, notably surface abundances and color-magnitude
diagrams of clusters, triggered the development of stellar evolution models for
rotating stars \citep[e.g.,][]{Maeder2000,Brott2011,Ekstrom2012}.  These usually
adopt the approximation of shellular rotation \citep{Zahn1992}, for which the
shape of the star is treated in the framework of the Roche model, with the polar
and equatorial radii differing maximally by a factor 1.5 at the critical
rotation rate $\Omega_{\rm crit}\equiv\sqrt{8GM/27R_{\rm p}^3}$ with $R_{\rm p}$
the polar radius of the star.

Rotation produces latitudinal dependence of the radiative flux, of the effective
temperature, of the stellar wind, etc. 
A basic assumption relied upon to compute models of rotating
stars is that the angular momentum lost from the outer envelope at each
time step due to a stellar wind, implies a change in the angular velocity
distribution inside the star.  Moreover, as the core contracts and the envelope
expands during the evolution, the rotation profile changes.  This
introduces a myriad of flows and instabilities causing chemicals to mix and
angular momentum to get transported in the radiative zones of stars.  Extensive
theory has been developed to describe
rotationally-induced instabilities and their accompanying mixing processes, both
in low-mass stars \citep[e.g.,][]{Chaboyer1995} and in intermediate- and
high-mass stars \citep[e.g.,][]{Talon1997,Pinsonneault1997,Heger2000}.  These 
processes are sometimes implemented with free
parameters in evolution codes, e.g., representing the level of chemical mixing
and/or 
angular momentum transport.  The transport processes are subject to considerable
uncertainties because they cannot be properly calibrated by classical
observations and few have been tested against simulations. With the advent of
asteroseismology, the theory can finally be evaluated. As an example, we
confront below the theoretical 
prediction that $\Omega_{\rm core}/\Omega_{\rm surf}\in [1,8]$ for
main-sequence stars with $1.7\simkl M\simkl 15\,$M$_\odot$
\citep[e.g.][Figure\,7]{Georgy2013}.

\subsection{\label{classical}Classical Observational Constraints 
for Stellar Interiors}

\begin{marginnote}[]
\entry{$\Omega_{\rm surf}$}{The angular surface rotation frequency of a star}
\entry{$\Omega_{\rm env}$}{The average angular rotation frequency in the stellar
  envelope}
\entry{$\Omega_{\rm core}$}{The average angular rotation frequency in the core
  region of the star}
\entry{$\Omega_{\rm crit}$}{The Roche critical angular rotation frequency of the
  star defined as $\sqrt{8GM/27 R_{\rm p}^3}$ with $R_{\rm p}$ the polar radius}
\end{marginnote}

High-resolution high signal-to-noise spectroscopy is a major classical
observational method to evaluate stellar evolution models. Spectroscopy allows
estimation of the effective temperature ($T_{\rm eff}$), gravity ($\log\,g$),
projected surface rotation $v\sin\,i\equiv \Omega_{\rm surf} R\,\sin\,i$ with
$R$ the stellar radius, and photospheric abundances of stars.  The relative
precisions of these spectroscopic diagnostics, which are widely available for
stars in the Milky Way and Magellanic Clouds, are compared with other
diagnostics for stellar interiors in Table\,\ref{precisions}.  Estimation of
$(T_{\rm eff},\log\,g,v\sin\,i)$ depends on spectrum normalization, particularly
of hydrogen lines, and suffers from degeneracies.  Despite this, relative
uncertainties for $T_{\rm eff}$ may reach the level of 1\% for low-mass stars,
but one usually cannot do better than 5\% for high-mass stars due to the limited
number of spectral lines.  Relative uncertainties for $\log\,$ are worse due to
the degeneracies and are therefore not included in Table\,\ref{precisions}.
Estimation of $v\sin\,i$ only offers indirect information due to the unknowns
$\sin\,i$ and $R$. Moreover, $v\sin\,i$ is also subject to uncertainties
caused by other spectral line broadening phenomena
\cite[e.g.][]{SimonDiazHerrero2014,Aerts2014a}. Photospheric abundances deliver
a powerful spectroscopic constraint and have been used extensively to
evaluate stellar evolution theory.  Low-mass stars in the solar neighborhood can
be analysed differentially with respect to the Sun, such that systematic
uncertainties cancel out. For these, the relative errors of [Fe/H] may be as low
as 1 to 2\% \citep[e.g.,][]{Melendez2014, Nissen2015}.  In the case of high-mass
stars, the relative errors for [C/H], [N/H], [O/H], and [Fe/H] are typically at
least 5 to 10\% \citep{Morel2008,Przybilla2013,Martins2015}.  A
positive correlation between the surface nitrogen abundance and $v\sin\,i$ has
been established for about half of the OB-type stars \citep[e.g.,][for a recent
discussion]{Dufton2018}.  However, the nitrogen abundance turned out to be
uncorrelated with the measured surface rotation frequency, $\Omega_{\rm surf}$,
and magnetic field for a sample of 68 slowly to moderately rotating B-type field
stars \citep{Aerts2014b}. Rather, evidence for a weak correlation between the
dominant oscillation mode frequency and the measured nitrogen abundance was
found, suggesting some sort of pulsational mixing in this sample.  Recent 2D
hydrodynamical simulations of convectively driven internal gravity waves (IGWs)
for a 3\,M$_\odot$ star indeed predict efficient particle mixing in radiative
layers \citep{RogersMcElwaine2017}.
\begin{marginnote}[]
\entry{IGWs}{Internal Gravity Waves}
\end{marginnote}

\begin{table}[h]
\tabcolsep3.0pt
\caption{Observational diagnostics used to calibrate  models of stellar  interiors
  and their optimal relative precision. 
  Type of star stands for LM: low mass; IM:
  intermediate mass; HM: high mass; RG: red giant; all: all masses; 
IRFM/SED: Infra Red Flux Method or Spectral Energy Distribution;
EB/SB2:
  double-lined eclipsing binaries; PB1/PB2: binaries with one or two pulsating components.
  We indicate the diagnostic's  dependence on stellar  atmosphere (A) 
  and/or stellar interior models (I).}
\label{precisions}
\begin{center}
\begin{tabular}{rcccc}
\hline
Method & Type of star & Diagnostic & Precision & Model dependence \\
\hline
Spectral lines & LM & $T_{\rm eff}$, abundances & $\sim 1\%$ & A: medium \\
Spectral lines & IM & $T_{\rm eff}$, abundances & $\sim 2\%$ & 
A: medium\\
Spectral lines & HM & $T_{\rm eff}$, abundances & $\sim 5\%$ & A: medium \\
\hline
IRFM/SED$^{\rm a}$   &  LM, IM &  $T_{\rm eff}$ & $\sim 2\% $ &  A: low\\
\hline
RV \& light curves$^{\rm b}$ & EB/SB2 & 
$M$ & $\sim 1\%$ & none \\
RV \& light curves$^{\rm b}$ & EB/SB2 & 
$R$ & $\sim 3\%$ & A: low \\
\hline
Interferometry$^{\rm c}$  & all & $R$ & $\sim 3\%$ & A: low\\
\hline
Typical Gaia DR2$^{\rm d}$  & LM \& IM & $L$ & $\simkl 15\%$ & A: medium\\
Typical Gaia DR2$^{\rm d}$  & LM \& IM & $R$ & $\simkl 10\%$ & A: medium\\
\hline
Cluster (E)MST$^{\rm e}$ & all  & age  & $\sim 30\%$ & I: strong\\
\hline
Gyrochronology$^{\rm f}$  & LM & $\Omega_{\rm surf}$ & $\sim 10\%$ & 
none \\
Gyrochronology$^{\rm f}$  & LM & age & $\sim 20\%$ & I: medium \\
\hline
\hline
Coherent g~modes$^{\rm g}$ & IM & $\omega_{nlm}$ & $\sim 0.1\%$ & none\\
Coherent p~modes$^{\rm g}$ & IM & $\omega_{nlm}$ & $\sim 0.01\%$ & none\\
Damped p~modes$^{\rm g}$ & LM & $\omega_{nlm}$ & $\sim 0.001\%$ & none\\
Damped mixed modes$^{\rm g}$ & RG & $\omega_{nlm}$ &  $\sim  0.01\%$ & none\\
g-mode splittings$^{\rm g}$ & IM & $\Omega_{\rm core}$ & $\sim 0.1\%$ &
none\\
g-mode spacings$^{\rm g}$ & IM & $\Omega_{\rm core}$ & $\sim 5\%$ &
I: low\\
p-mode splittings$^{\rm g}$ & IM & $\Omega_{\rm env}$ & $\sim 30\%$ &
I: medium\\
p-mode splittings$^{\rm g}$ & LM & $\Omega_{\rm env}$ & $\sim 50\%$ &
I: medium\\
mixed-mode splittings$^{\rm g}$ & RG & $\Omega_{\rm core}$ & $\sim  1\%$
                                               & none\\
Phase Modulation \& RV$^{\rm g,h}$ & PB1/PB2 & $M, R$ & as EB/SB2 & none\\
\hline
\end{tabular}
\end{center}
\begin{tabnote}
  $^{\rm a}$ provided that a good estimate of the 
reddening and an absolute flux calibration are
  available; this method can also lead to a radius estimate at a level
  $\sim\!3\%$ when a high-precision parallax (at the level of Gaia DR3)
becomes available;
  $^{\rm b}$ modeling of double-lined eclipsing binaries (EB/SB2) based on
  radial velocities (RV) from spectroscopy and light curves from photometry,
  both covering the orbit of the binary \citep{Torres2010}; $^{\rm c}$ requires
  good calibration stars, as well as a high-precision parallax and
  $T_{\rm eff}$, limiting the application to bright stars; 
$^{\rm d}$ typical values covering 77 million stars \citep{Andrae2018};
$^{\rm e}$~isochrone
  fitting of (Extended) Main-Sequence Turnoffs, (E)MST, of clusters from
  combined multi-color photometry, spectroscopy, and astrometry
  \citep{BastianLardo2018}; $^{\rm f}$ from $\mu$mag precision time-series space
  photometry covering many cycles of the rotationally modulated light curve due
  to spots, delivering $\Omega_{\rm surf}$; $^{\rm g}$ requires long-duration
  $\mu$mag precision uninterrupted space photometric light curves, delivering
  the frequencies $\omega_{nlm}$ of tens of identified nonradial oscillation
  modes $(n,l,m)$, where we additionally note that 
combined $T_{\rm eff}$, Gaia DR2, and damped modes lead to radii with
  $\sim\!1\%$ precision for the best cases \citep{Zinn2018};
  $^{\rm h}$ the potential of this method is largest for p modes and depends on
  the nature of the pulsating binary (one or two pulsating components), 
where PB2/SB2 have similar potential as EB/SB2
  \citep[cf.,][for details and first applications]{Murphy2016}.
\end{tabnote}
\end{table}

Double-lined spectroscopic eclipsing binaries have long been known as
excellent calibrators for stellar models. Indeed, spectroscopic and (multicolor)
photometric data covering their orbit offer model-independent stellar masses at
the level of 1--3\% from the binary motion \citep[][for an extensive review
based on 95 objects, covering the mass range 0.2 to 27\,M$_\odot$]{Torres2010}.
{\it Kepler\/} space photometry revealed similar potential for non-eclipsing
double-lined binaries pulsating with coherent p~modes from the Phase Modulation
method \citep{Murphy2016}.  Double-lined eclipsing binaries have also been used
to estimate convective core overshooting, suggesting a mass-dependence
\citep{Claret2018}. However, core overshooting depends strongly on various
other stellar parameters and such correlations have not yet been accounted for in
binary modeling, while this should be done (Johnston et al., submitted).

Interferometric data offers good estimates of angular diameters,
translating into precise stellar radii to evaluate stellar models for the very
brightest stars. This technique offers a level of precision for $R$ of
$\sim\!3\%$ for nearby stars, provided that a good parallax is available
\citep[e.g.][]{Ligi2016}.  Two major limitations with this methods are 
the limited availability of good calibrators and uncertainties in the
limb darkening \citep[e.g.][]{White2018}. In addition to
interferometry of bright nearby stars, Gaia astrometry offers the potential to
derive accurate radii for millions of stars in the Milky Way, provided that a
good estimate of $T_{\rm eff}$ is available.  Gaia\,DR2
already led to radii at $\sim\!2\%$ precision for a number of (exoplanet
host) stars \citep[e.g.,][]{Fulton2018} and even $\sim\!1\%$ for asteroseismic targets
\citep{Zinn2018}. At the level of Gaia\,DR3 (to be
released in 2021), stellar radii with a precision of $\sim\!1\%$ are anticipated
as the norm for millions of single and binary stars.

Isochrone fitting of color-magnitude diagrams of stellar clusters has been used
extensively to test models of stellar interiors, by deducing stellar ages from
the observed turnoff point.  Large observing
campaigns with the Hubble Space Telescope led to systematic detections of
multiple main sequences in globular clusters covering masses from 0.1 to
1.8\,M$_\odot$ and ages from 2 to 15 Gyr \citep[e.g.,][]{Piotto2002,Dotter2007}.
These are interpreted as due to different populations of stars with abundance
variations indicative of nuclear processing by the CNO cycle \citep[e.g.,][for
early discoveries and an extensive review,
respectively]{Piotto2007,BastianLardo2018}.  On the other hand, open clusters
with intermediate- and high-mass stars in the Milky Way and Magellanic Clouds
reveal extended main sequence turnoffs and/or split main sequences.
Interpretation of these in terms of stellar models with rotation leads to ages
in agreement with the lithium depletion method.
It remains unclear what level of the extended turnoffs is due to rotation (or
other phenomena) and what level is due to a spread in age
\citep[e.g.,][]{Gossage2018,Bastian2018}.  Interpretation of the observed
extended turnoffs as solely due to age leads to about 30\% uncertainty for
stellar aging.  The luminosity ($L$) estimates deduced from Gaia DR2 data led to
spectacular improvements of HR diagrams of clusters
\citep{Babusiaux2018} and Gaia DR3 data is anticipated to bring
more potential for the improvement of stellar interiors from fitting of the
overall cluster morphologies.

Another way to derive stellar age as an observational test of stellar models is
by the method of gyrochronology.  Already prior to high-precision space
photometry, it was realized that the presence of an envelope dynamo in low-mass
stars offers a good way of aging them \citep{Barnes2003}.  Gyrochronology is
based on the stellar spin-down due to magnetic breaking and uses the surface
rotation rate as a clock, adopting the so-called Skumanich relation between
angular momentum loss and rotation rate, $dJ/dt \propto \Omega_{\rm surf}^3$
\citep{Skumanich1972}. This has meanwhile been extensively applied to {\it
  Kepler\/} data of low-mass (exoplanet-hosting) field and cluster stars
\citep{Garcia2014,Meibom2015}. Latitudinal surface differential rotation limits
gyrochonology for field stars to the level of $\sim\!2\,$Gyr, i.e., some 20\%
relative precision \citep{Epstein2014}. Moreover, the occurrence of anomalously
rapid rotation in old field stars hints at weakened magnetic breaking at evolved
stages of low-mass stars \citep{vanSaders2016}. This implies that more complex
models than a simple empirical relation between the rotation rate and stellar
age are required for precise stellar aging from gyrochronology of low-mass
stars.  Such models can be designed by calibrating gyrochronology with
asteroseismically determined ages \citep{Garcia2014,Ceillier2016}. This brings
us to the recent breakthroughs in stellar modeling based on asteroseismology,
keeping in mind the broad availability of classical diagnostics to evaluate
stellar interiors summarized in Table\,\ref{precisions}.

\section{\label{Astero-age}THE NEW AGE OF ASTEROSEISMOLOGY}

\subsection{Asteroseismology in a Nutshell}

Stellar oscillations (aka starquakes) offer a direct probe of stellar interiors.
Asteroseismology, the interpretation of detected identified oscillation modes
\cite[][]{Aerts2010}, offers a unique way to evaluate and calibrate stellar
interiors. It uses the measured high-precision frequencies of stellar
oscillations as revealed by time-domain astronomy as an observational diagnostic
of the stellar interior to evaluate stellar models, in addition to classical
observables (Table\,\ref{precisions}).  Asteroseismology is the
``champion'' in the observational probing of stellar interiors.

\begin{marginnote}[]
\entry{Asteroseismology}{the interpretation of detected frequencies of identified 
oscillation modes with radial order, degree and azimuthal order, $(n,l,m)$, in
terms of the physical properties of the star's interior}
\end{marginnote}

Asteroseismology follows the foundations laid out for the interpretation of the
solar oscillations by means of helioseismology \citep{JCD2002}. The application
of this method to distant stars was made possible thanks to uninterrupted
high-precision photometric time-series data assembled by recent space
missions. In this review, we mainly rely on light curves assembled by the NASA
{\it Kepler\/} telescope \citep{Koch2010} and its refurbished K2
version \citep{Howell2014}. These data have a precision of $\sim\!\mu\,$mag,
cover up to four years and three months, respectively, and  
offered the long awaited evaluation of the theory of stellar
interiors. It turns out that shortcomings in the 
theory of angular momentum transport 
already occur on the main sequence, impacting all subsequent phases 
\citep[e.g.][]{Eggenberger2017,Townsend2018,Tayar2018}. As such,
time domain space photometry assembled during the last decade is a game changer
in the calibration of stellar structure and evolution theories.

\begin{textbox}[h!]\section{The plethora of coherent and damped 
linear oscillation modes inside  stars}
Whenever a spherically symmetric star in hydrostatic equilibrium gets perturbed,
it may become unstable.  From a theoretical viewpoint, small disturbances of the
equilibrium configuration are treated in a linear approach. As such, the
reactions of the acting forces in the equation of motion are investigated to
understand and interpret stellar (in)stability
\citep{Ledoux1941,Unno1989}.  The instabilities may give rise to
detectable self-driven coherent oscillation modes with long lifetimes (thousands
to millions of years) whenever their growth rate is positive and when the
thermal time scale of the excitation layer in the stellar envelope is longer
than the period of the oscillation modes \citep[e.g.,][]{Cox1980,Pamyatnykh1999}.  But
even when modes are damped, they may still be excited to observable amplitudes
by stochastic forcing. This is the case for stars with an outer convective
envelope such as the Sun \citep{JCD2002} and red giants
\citep{ChaplinMiglio2013}, and gives rise to oscillation modes with lifetimes of
days to weeks.

Perturbation of the linearized stellar structure equations leads to the
properties of the oscillations of a particular star, among which its mode
frequencies.  The modes are categorized according to the dominant restoring
force.  Pressure (or acoustic), gravity, inertial, and magnetic (or Alfv\'en)
oscillations occur when the dominant restoring force is the pressure force,
gravity (buoyancy), the Coriolis force, and the Lorentz force, respectively. 
When stars are hosting close planetary or stellar companions, tidal
forces may trigger these different types of modes; in this case they are called tidal
oscillations. In practice, the pressure force and gravity are always active,
while the other forces may be of minor importance. Hence, one has
introduced the specific terminology of p (for pressure) and g (for gravity)
modes.  In reality, the forces act together to restore the equilibrium of the
star and one may encounter the situation where two (or more) among them are of
similar importance. In that case, the distinction of the modes in terms of
frequency may become less clear and the terminology is adapted accordingly,
e.g., gravito-inertial modes, magneto-acoustic modes, etc.
\end{textbox}

Asteroseismic probing of stellar interiors is achieved by interpreting the
properties of detected nonradial oscillation modes, notably their frequency.
Gravity modes allow one to probe the deepest layers of stars, where chemical
gradients are built up during stellar evolution. On the other hand, the p~modes
probe the outer envelope of stars. The oscillation modes can often be
represented by simple waves, whose angular frequencies $\omega$ fulfill
dispersion relations \citep[][for a summary of mathematical
treatments]{Smeyers2010}. As an example, p~modes correspond to acoustic waves
and g~modes to IGW.  For acoustic waves, the dispersion relation involves the
sound speed of the gas, while for gravity waves it is the buoyancy (aka
Brunt-V\"ais\"al\"a) frequency, $N$. While pressure waves can propagate in both
convective and radiative regions, gravity waves can only propagate in
convectively stable radiative layers.  Just as oscillation modes can be coherent
or damped in nature, the corresponding waves can have a standing or traveling
wave nature in their propagation cavities inside the star. A broad spectrum of
both damped and standing IGW can be generated from partial ionization layers in
radiative envelopes and stochastic forcing at convective-radiative
interfaces. While the traveling (damped) IGW cannot be used for stellar modeling
like their coherent counterparts, unless they coincide with eigenmodes of the
star, they do transport angular momentum.

\begin{marginnote}[]
\entry{$\omega$}{Angular frequency of an oscillation mode}
\entry{$N$}{Buoyancy (or Brunt-V\"ais\"al\"a) frequency}
\end{marginnote}

For an extensive discussion on the theory of nonradial oscillation modes, their
excitation, their relation to waves and their propagation, we refer to the
monographs by \citet{Unno1989} and \citet{Aerts2010}. The latter book also
offers methodology for the frequency analysis and identification of nonradial
oscillation modes, in great detail. Here, we limit the discussion to the bare
minimum necessary to understand the exploitation of detected oscillation modes
and IGW in terms of the stellar interior, with emphasis on rotation. For modern
descriptions of coherent mode excitation, we refer to
\citet{Dupret2005,Bouabid2013,Szewczuk2017}. IGW generation by convective cores,
convective envelopes or thin convection zones due to opacity bumps in radiative
envelopes is extensively discussed in
\citet{Rogers2013,Talon2008,Cantiello2009}, respectively.

Unlike ground-based time-series data with low duty cycle, high-cadence
uninterrupted space photometry does not lead to daily aliasing confusion in the
derivation of the oscillation frequencies. As such, the frequency
uncertainty is dominated by the frequency resolution of the data set. For
coherent modes, this resolution is $\sim\!1/T$, with $T$ the
total time base of the time-series data. This is 0.00068\,d$^{-1}$
(0.0079$\mu$Hz) for the 4-year nominal {\it Kepler\/} data.  For damped modes,
the resolution is $\sim\!1/\sqrt{T}$.  Aside from the resolution of the
data, the frequency error of each detected mode depends on its
amplitude and the noise properties of the data in the relevant frequency regime
\citep[][Chapter\,5]{Aerts2010}.  Damped modes have similar amplitudes and allow
their frequencies to be deduced from peak bagging \citep[e.g.,][]{Lund2017}.
Coherent modes, on the other hand, may have mode amplitudes differing six orders
of magnitude and their frequencies must therefore be deduced from a prewhitening
procedure \citep[e.g.,][Chapter\,5]{Aerts2010}. Iterative prewhitening
inherently introduces uncertainties of order of the frequency resolution of the
data.  The nominal 4-year {\it Kepler\/} data lead to a relative frequency
precision better than 0.1\% for coherent g~modes and 0.01\% for coherent p modes
in intermediate- and high-mass stars.  Damped p~modes of low-mass stars can be
measured with a typical relative precision better than 0.001\% for dwarfs
\citep{Lund2017} and 0.01\% for red giants \citep{HekkerJCD2017,Mosser2018}.

Exploitation of such ultra-precise asteroseismic information has brought a
wealth of information on stellar interior processes that is inaccessible from
classical data. We dedicate this review to the asteroseismic inference of
interior rotation as it constrains internal angular momentum transport.  As will
be explained below, the asteroseismic estimation of the rotation rate as a
function of depth $r$ inside the star, $\Omega(r)$, can be fully or quasi
model-independent. Which of the two it is, depends on the kind of star, the
nature of the detected oscillations, and the level of stellar rotation, as
explained in Section\,\ref{section-rotation}.  First we discuss how an
asteroseismically calibrated model of a star can be sought from its identified modes,
before moving on to estimation of the interior rotation.

\begin{textbox}[b!]\section{Spherical harmonics and rotational splitting 
in asteroseismology}
Following the linear theory of stellar oscillations applied to a spherically
symmetric equilibrium star, the displacement vector of a non-radial oscillation mode of
degree $l$ and azimuthal order $m$ is given by
$\xi (r, \theta, \phi, t) = [(\xirnl e_{r} + \xihnl \nabla_{\rm h})
Y_l^m(\theta,\phi)] \exp (- {\rm i} \omega t )$, with $\theta$ the angle
measured from the polar axis, $\phi$ the longitude, and $\omega$ the angular
mode frequency.  In the absence of rotation, the modes are called zonal
(for which $m=0$); these reveal $l$ surface nodal lines.

  In the presence of rotation, the rotation axis of the star is usually chosen
  as the axis of the spherical polar coordinate system to describe the modes. In
  this case, $|m|$ ($\leq l$) of the surface nodal lines are lines of longitude.  A
  distinction is made between prograde $m>0$ and retrograde $m<0$ modes,
  corresponding with waves travelling with and against the rotation,
  respectively.  The Coriolis acceleration due to the rotation of the star causes
  rotational splitting of the mode frequencies into $2l+1$ components called
  multiplets.  The level of splitting among the multiplet components is
  determined by the rotation frequency $\Omega$ of the star, where larger
  splitting is due to faster rotation.

  The radial order of the mode, $n$, representing the number of nodes of $\xir$,
  cannot be deduced from surface observables. Rather, it can be estimated from
  comparison between observed and theoretically predicted frequencies after
  identification of $(l,m)$. We refer to the monograph \citet{Aerts2010} for
  detailed methodology to achieve this.
\end{textbox}

\subsection{\label{forward-low}Asteroseismic Modeling of Low-mass Stars}

Asteroseismic modeling relies on 1D spherically symmetric equilibrium models and
computations of their 3D eigenmodes and eigenfrequencies by perturbing these 1D
models. For low-mass main-sequence stars, such modeling is based on damped
p~modes excited by the envelope convection, which reveal themselves with their
individual frequencies superposed on a power excess due to granulation
\citep[e.g.,][]{Kallinger2014}.  The p~modes have short periods of a few minutes
and mainly probe the outer envelope.  Given the slow rotation of such stars,
their zonal ($m=0$) p~modes can be exploited while neglecting the Coriolis and
centrifugal forces. Also the Lorentz force has a negligible effect on the
oscillations compared to the pressure force and gravity.  The asteroseismic
modeling of the zonal-mode frequencies is often based on the same input physics
as for the 1D solar model derived from helioseismology \citep{JCD2002}.

Low-degree zonal p~modes in low-mass stars reveal a characteristic frequency
spacing related to the mean density of the star, while the
frequency at maximum power is related to the acoustic cut-off frequency  
depending on $M$, $R$, and
$T_{\rm eff}$.  A spectroscopic measurement of $T_{\rm eff}$ thus allows to
derive $M$ and $R$ with respect to those of the Sun. This
potential was predicted long before the space era of asteroseismology in the
seminal paper by \citet{KjeldsenBedding1995} and has by now been applied to
hundreds of low-mass
dwarfs and subgiants \citep{Chaplin2014}.  Once $M$ and $R$ have been derived,
a model-dependent stellar age follows. This procedure leads to relative
precisions of $\sim\!2\%$ in radius, $\sim\!4\%$ in mass, and $\sim\!10\%$ in age
for stars that were monitored during the 4-year nominal {\it Kepler\/} mission
and that have similar metalicity and interior physics to the Sun
\citep{SilvaAguirre2017}.

Evolved low- and intermediate-mass stars offer, in addition to their p~modes,
the opportunity to exploit mixed dipole ($l=1$) modes.  These have a p-mode
character in the envelope and a g-mode character in the deep interior.  In
contrast to p~modes, g~modes have characteristic period spacings determined by
the buoyancy frequency $N$.  Such spacings were detected from ground-based
photometry for white dwarfs long before space asteroseismology \citep[e.g.,][for
an overview]{Kawaler1999}. Red giants reveal mixed modes with gravity-dominated
or pressure-dominated character. Evolved stars thus offer both frequency
spacings from their $l=0,2$ p~modes and period spacings from their
gravity-dominated mixed $l=1$ modes.  The potential of mixed modes was studied
theoretically by \citet{Dupret2009} and discovered in {\it Kepler\/} data by
\citet{Beck2011} and \citet{Bedding2011}. The overall probing power of mixed
dipole modes allows one to pinpoint the nuclear burning phase of red giants --
an assessment that cannot be done from surface measurements
\citep{Bedding2011,Mosser2014} -- but also how the star's core rotates
\cite[][cf.\
Section\,\ref{section-rotation}]{Beck2012,Mosser2012,Deheuvels2014}.  Extensive
reviews on 1D red giant modeling based on frequency and period spacings of their
damped modes are available in \citet[][]{ChaplinMiglio2013} and
\citet{HekkerJCD2017}.

\subsection{\label{forward-high}Asteroseismic Modeling of Intermediate- 
and High-mass Stars}

The overall mixing profile beyond the Schwarzschild boundary of the convective
core in intermediate-mass and high-mass stars, $D_{\rm mix}(r\!>\!r_{\rm cc})$, is
affected by several phenomena, such as convective overshooting/penetration,
semi-convection, rotationally- and magnetically-induced instabilities,
thermohaline mixing, etc.  Theoretically predicted instabilities in the
radiatively stratified envelope due to rotation or magnetism
lead to discontinuities in
$D_{\rm mix}(r\!>\!r_{\rm cc})$. These often remain invisible in evolutionary
tracks, but they affect 3D oscillation mode computations. The latter hence offer
a powerful way to calibrate the (free parameters assigned to) mixing and angular
momentum transport due to instabilities.  In order to achieve that,
asteroseismic modeling of an observed star is best done by relying on 1D
non-rotating non-magnetic equilibrium models and by computing their mode
frequencies. The computation of these
3D oscillations must take into account the accelerations due to rotation in the perturbed equation of motion. This allows to select the best
equilibrium model for the stellar interior and to estimate its free
parameters, such as mass, metalicity, and age. Deviations between the predicted
and measured oscillation frequencies then reveal shortcomings in the input
physics of the equilibrium models and offer a guide to improve it up to the
level of precision of the identified oscillation frequencies.

Early asteroseismic modeling of the high-mass B3V star HD\,129929 from six
identified low-order $l=0, 1, 2$ modes detected in ground-based data revealed
core convective penetration with $\alpha_{\rm ov}=0.15$
\citep{Aerts2003,Dupret2004}.  In addition, the rotationally split modes led to
$\Omega_{\rm core}/\Omega_{\rm env}\!\sim\!4$.  Following this initial study,
major ground-based multi-site campaigns lasting many months and involving tens
of astronomers were organised \citep[][for some
examples]{Handler2006,Briquet2012}. These led to
$\alpha_{\rm ov}\in [0,0.45]\,$H$_p$ assuming convective penetration.
So far, space missions have observed few high-mass stars and none with suitable
identified modes to improve asteroseismic modeling achieved from the ground.

Intermediate-mass stars offer better opportunities, as they reveal tens of
low-degree, high-order g~modes, with periods of half to a few days.  The first
detection of g-mode period spacings in main sequence stars came from the CoRoT
mission \citep{Degroote2010} and revealed optimal probing power in the near-core
region of stars with a convective core.  Such g-mode spacings have meanwhile
been detected for a whole range of rotation rates, from very slow to 80\%
critical (cf.\ Section\,\ref{section-rotation}).  Since the g-mode periods may
be of similar order to the rotation period, typically 1 to 3 days, the Coriolis
force is a mandatory ingredient in the 3D oscillation computations to interpret
the observed period spacing patterns of g~modes, as demonstrated by
\citet{VanReeth2015,VanReeth2016,Ouazzani2017} for F-type pulsators and by
\citet{Moravveji2016,Papics2017,Saio2018a,Szewczuk2018} for B-type pulsators.
Unlike for low-mass stars, asteroseismic modeling based on g~modes in main
sequence stars therefore starts with estimation of the near-core rotation rate
$\Omega_{\rm core}$, as explained in Section\,\ref{section-rotation}.

\begin{marginnote}[]
  \entry{TAR}{Traditional Approximation of Rotation, ignoring the horizontal
    component of the rotation vector in the equation of motion for a spherically
    symmetric star}
\end{marginnote}

Another complication compared to low-mass stars is the need to include physical
ingredients that do not occur in the Sun, such as convective
overshooting/penetration from the core instead of the
envelope. \citet{Aerts2018a} and Johnston et al.\ (submitted) developed a
methodological framework to perform asteroseismic modeling of single and binary
stars with a convective core, based on tens of identified g~modes of consecutive
radial order ($n$ typically between -5 to -50), to estimate
$(M,X,Z,\Omega_{\rm core},D_{\rm mix}(r\!>\!r_{\rm cc}),X_c)$, where $(X,Z)$ are the
relative mass fractions of hydrogen and metals such that $X+Y+Z=1$ (with $Y$ the
helium mass fraction), and $X_c$ is the hydrogen mass fraction in the convective
core (a proxy of the age during the main sequence).  Their method adopts the
so-called Traditional Approximation of rotation (TAR) for the treatment of the
Coriolis force \citep[e.g.,][]{Townsend2003b,Townsend2003a}.
Estimation of $D_{\rm mix}(r\!>\!r_{\rm cc})$ by g~modes is particularly powerful in
the core overshoot zone and allows one to derive the core mass of the star
\citep{Moravveji2015,Moravveji2016}.  Further, the modeling required
$D_{\rm mix}(r\!>\!r_{\rm cc})\neq\,0$ in order to explain the observed g-mode
trapping properties, with measured levels of
$D_{\rm mix}\in [1,1\,000]$\,cm$^2$\,s$^{-1}$. This is smaller than predictions
by the theory of rotationally- or magnetically-induced instabilities and
provides a new way to calibrate and guide that theory
\citep[e.g.][]{Mathis2013LNP}, as well as simulations
\citep{RogersMcElwaine2017,Pedersen2018}.

Element and angular momentum transport in stars are intimitely related because
they result from the same processes.  Often, they are decoupled from each
other in stellar evolution codes, where each of the transport phenomena is given
its own free parameter because of a lack of transport theories based on first
principles.  From here on, we focus on angular momentum transport, given
that the asteroseismic measurement of $\Omega_{\rm core}$ is (quasi-)independent
of 1D equilibrium models, while estimation of the mixing profile inside
the star does depend on it (cf.\,Table\,\ref{precisions}).  We refer to
\citet{Salaris2017} for a review of the theory of element transport in stars
prior to the asteroseismic estimates available today.

\subsection{\label{section-rotation}Asteroseismic Derivation 
of Interior Rotation}

The derivation of the interior rotation rates from asteroseismology cannot be
done for all pulsators, because it requires the detection and identification of
specific oscillation modes. Detection of rotationally split mixed or g~modes
is required to derive $\Omega_{\rm core}$, while split p~modes lead to
$\Omega_{\rm env}$.  The approach to follow depends on the ratio of $\Omega$ and
$\omega$ and is therefore different for p and g~modes in different types of
stars. Whenever $\Omega/\omega\simkl 25\%$, one can rely on a perturbative
approach to assess the effects of the Coriolis and centrifugal forces on the
stellar oscillations. This is typically the case for p~modes in most stars, for
p and mixed modes in red giants, and for g~modes in subdwarfs and white
dwarfs. On the other hand, $\Omega/\omega\simgr 25\%$ requires a
non-perturbative treatment of the Coriolis and/or centrifugal forces. Regarding
the Coriolis force, this is conveniently done by the TAR for g~modes in
intermediate- and high-mass stars in the case where $2\Omega\!<\!\!<\!N$.

\subsubsection{From a Perturbative Approach}

Both the Coriolis and centifugal forces change the mode frequencies.  Following
a first-order perturbative approach in the computation of the oscillation
frequencies is fine when the centrifugal force ($\sim\!\Omega^2$) and hence the
deformation of the star can be ignored, and when $\Omega <\!\!< \omega$.  Let us
assume that this is the case and that the rotation rate only changes with depth
$r$ inside the star (and not with latitude). Under these
assumptions, the mode frequency $\omega_{nl}$ belonging to eigenvector
$\xi_{n l} $ $=( \xirnl , \xihnl )$ in the non-rotating
case gets split into $2l+1$ frequency multiplet components 
due to the Coriolis acceleration.
Moreover, each of those components gets shifted due to
the Doppler effect according to the rotation between a coordinate system
corotating with the star and the inertial coordinate
system of the observer. The frequencies of the multiplet
in the interial frame then become
\begin{equation}
\omega_{nlm}  = \omega_{nl}  + \ m\ (1-C_{nl}) \int_0^R K_{nl} (r) 
\Omega (r) \dd r  \; , 
\label{LedouxSplitting}
\end{equation}
where
\begin{equation}
K_{nl} (r) =
{ \left( {\xi_r}^2 +  [l(l+1)] \xih^2
 - 2 {\xi_r} \xih - \xih^2 \right) r^2 \rho   \over \int_0^R \left( {\xi_r}^2 + 
[l(l+1)] \xih^2 - 2 {\xi_r} \xih - \xih^2 \right) r^2 \rho \dd r} \; ,
\label{kernel}
\end{equation}
is called the unimodal rotational kernel and
\begin{equation}
C_{nl}  = 
{\int_0^R \left( 2 {\xi_r} \xih + \xih^2 \right) r^2 \rho \dd r  \over 
\int_0^R \left( {\xi_r}^2  + [l(l+1)] \xih^2 \right)  r^2 \rho \dd r} \; 
\label{Cnl}
\end{equation}
the so-called Ledoux constant \citep{Ledoux1951}.
Both $C_{nl}$ and
$K_{nl} (r)$ depend 
on the 1D equilibrium model of the star. However, as outlined in
detail in \citet[][Chapter\,3, Figure\,3.38]{Aerts2010}, $C_{nl}\simeq 0$
for high-order or high-degree p~modes.
On the other hand, the
displacement is dominantly horizontal for high-order g~modes and the terms
containing ${\xi_r}$ in Eqs\,(\ref{kernel}) and (\ref{Cnl}) can be neglected, leading to
\begin{equation}
C_{nl} \simeq {1 \over [l(l+1)]} \; ,
\end{equation}
which is independent of the equilibrium model.
\begin{figure}
\begin{center}
\rotatebox{270}{\resizebox{4.cm}{!}{\includegraphics{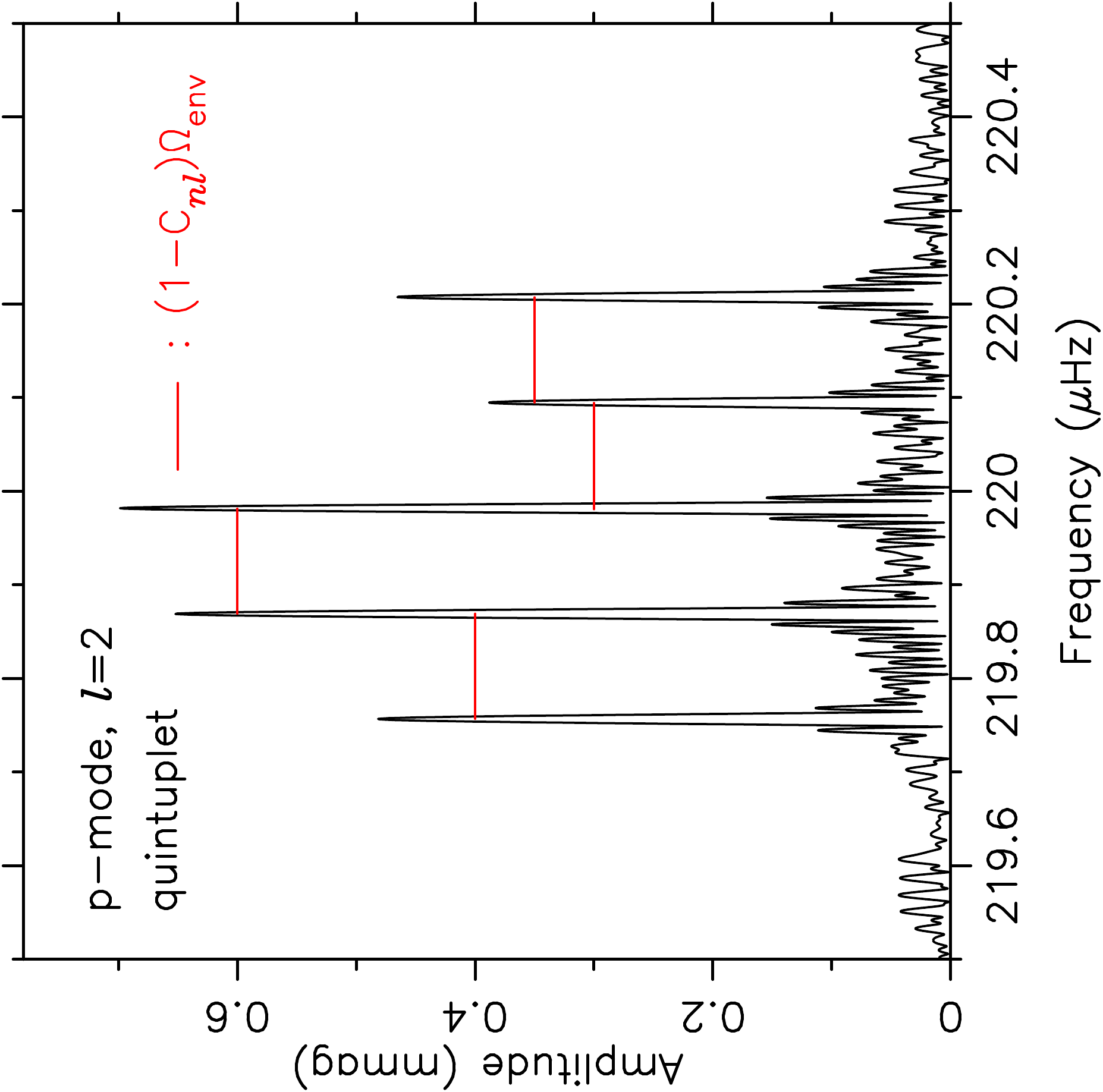}}}\hspace{0.5cm}
\rotatebox{270}{\resizebox{4.cm}{!}{\includegraphics{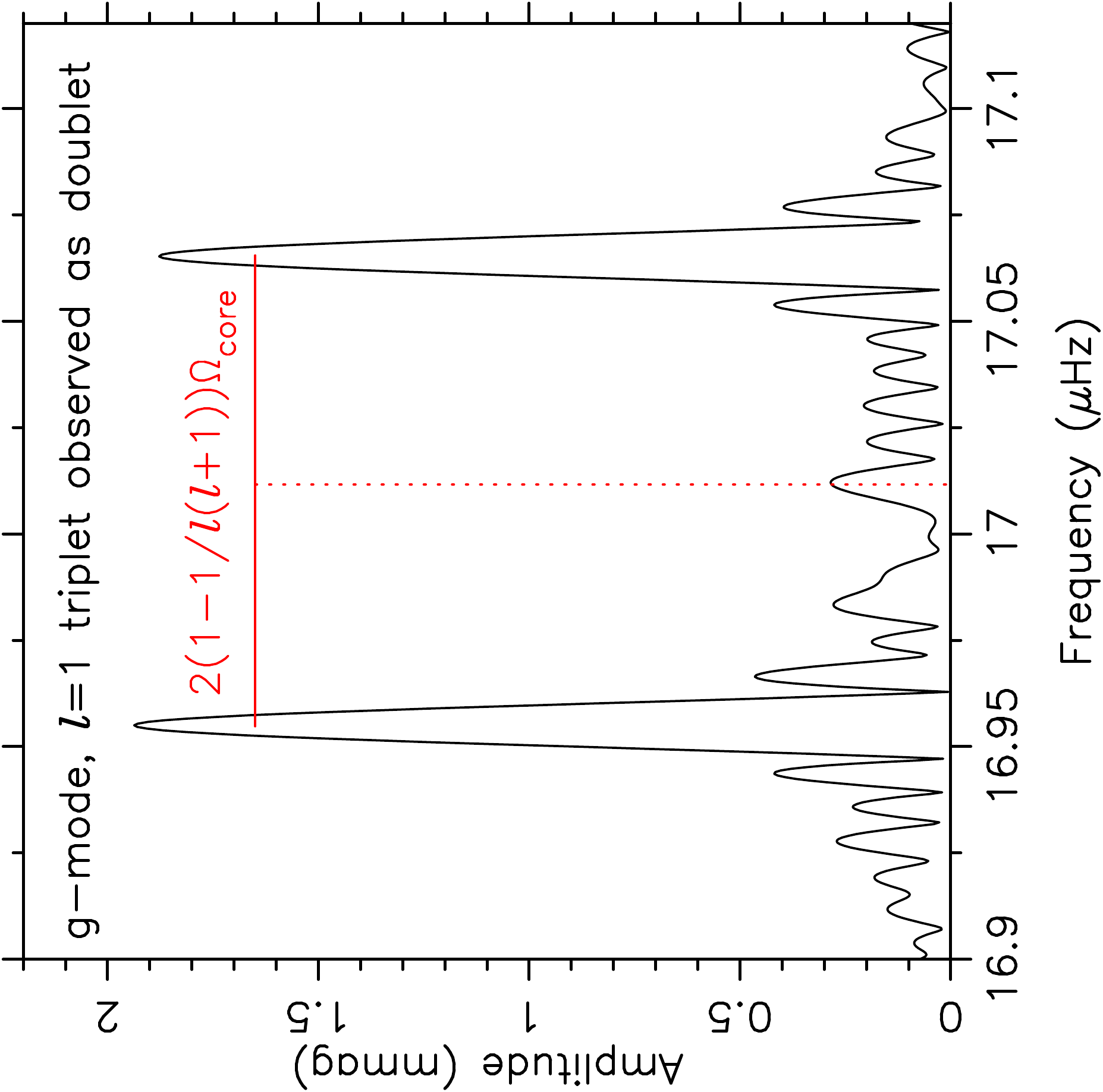}}}\\[1.cm]
\caption{\label{fig-splitting} Rotational splitting of a quadrupole p mode
  (left) and a dipole g-mode (right) of the intermediate-mass star
  KIC\,11145123. The red dotted line indicates the position of the $(l,m)=(1,0)$ g-mode frequency. Figure produced from data in \citet{Kurtz2014}.}
\end{center}
\end{figure}
In the approximation of uniform rotation, the
transformation between the corotating frame and the observer's inertial
frame implies a frequency shift of $m\Omega$ and one obtains
\begin{equation}
\omega_{nlm} = \omega_{nl} + m (1 -\  C_{nl}) \Omega \; .
\end{equation}
In this case the effect of rotation on the oscillation mode frequencies is
completely determined by $C_{nl}$ and does not depend on the rotational kernels.
For high-order or high-degree p~modes, for which $C_{nl} \simeq 0$, the measured
rotational splitting between adjacent $m$-values in a multiplet is then given by
the average rotation rate in the stellar envelope, $\Omega_{\rm env}$, in a
model-independent way.  For low-order p~modes, the derivation of
$\Omega_{\rm env}$ depends on the 1D equilibrium model through
$C_{nl}$. Depending on the type of star and its evolutionary state, typical
values are $C_{nl} \in [0,0.5]$ so the derivation of $\Omega_{\rm env}$
maximally implies a relative uncertainty of 50\% for $\Omega_{\rm env}$. Often,
however, the model dependence of $C_{nl}$ is negligible, leading to a
high-precision quasi model-independent estimate of $\Omega_{\rm env}$. This is
the case for the {\it Kepler\/} target KIC\,11145123, whose p-mode splitting is
illustrated in the left panel of Figure\,\ref{fig-splitting}.  For high-order
g~modes, $C_{nl}$ is fully determined by the mode degree, hence their measured
splitting offers a model-independent estimate of $\Omega_{\rm core}$. In
particular, the rotational splitting measured from adjacent frequency peaks in a
triplet of g~modes equals half the rotation rate. Aside from p~modes,
KIC\,11145123 also shows rotationally split g~modes. Given the almost equator-on
view upon the stellar rotation axis, its dipole g~modes are detected as doublets
since the zonal modes almost cancel out in the line-of-sight. This is
illustrated in the right panel of Figure\,\ref{fig-splitting}, where the
splitting between the two doublet components offers a direct and
model-independent measurement of $\Omega_{\rm core}$ \citep{Kurtz2014}.

In the general case of differential rotation, the measured rotational splitting
is determined by the eigenvector of the mode, $\xi_{n l}$, as well as by the
shape of $\Omega (r)$ and the rotational kernels, which depend on the 1D
equilibrium model.  In that case, the definition of the corotating frame is not
obvious, but a kernel-weighted average of $\Omega(r)$ is a logical choice to
compute the Doppler shift towards an inertial frame. As such, one obtains a
kernel-weighted average of $\Omega_{\rm env}$ from p-mode multiplets and a
kernel-weighted average of $\Omega_{\rm core}$ from g~modes or from mixed-mode
multiplets as detected in {\it Kepler\/} data of red giants
\citep{Beck2012,Mosser2012}.

Perturbation theories up to higher order in $\Omega$ have been developed, e.g.,
\citet[][2nd order, p~modes]{Saio1981}, \citet[][2nd order, p and g~modes
including magnetism]{Dziembowski1992}, \citet[][2nd order,
g~modes]{LeeBaraffe1995}, \citet[][3rd order, p~modes]{Soufi1998,Daszynska2002},
\citet[][2nd order, p, g, and mixed modes]{Suarez2006}. These theories are based
on various assumptions regarding the Coriolis, centrifugal, Lorentz, and tidal
forces, and for the ratios of $\Omega$, $\omega$, the Alfv\'en frequency
$\omega_{\rm A}$, the orbital frequency of a binary in the oscillation equations
and also on the way the deformation of the star $\sim\!\Omega^2$ is treated in
the computation of the equilibrium model. Those higher-order
perturbation theories leads to particular splittings and shifts in the
frequencies of the oscillations compared to $\omega_{nl}$ of the non-rotating
case.  The theories have in common that they lead to coupling of particular
eigenmodes of different $l$ and $m$. The period spacing and mode trapping
properties of these various perturbation theories may deviate appreciably from
those of non-perturbative theories \citep[cf.\ Figure\,6 in][for a concrete
comparison between TAR and a 2nd order theory for g~modes]{Saio2018b}.

\subsubsection{Based on the Traditional Approximation}

The perturbative approach is no longer appropriate when $\omega$ is of the same
order of, or smaller than, $\Omega$. The reason is that the Coriolis
acceleration is no longer small compared to the acceleration term in $\omega^2$
in the equation of motion.  As outlined by \citet{LeeSaio1987}, the oscillation
equations are appreciably simplified when the tangential component of the
rotation vector is ignored. This so-called TAR, commonly used in geophysics
\citep{Eckart1960} and in studies of neutron stars \citep{Bildsten1996}, 
allows modes to be computed from the Laplace tidal
equations and offers an excellent approximation for g~modes in intermediate- and
high-mass main sequence stars \citep[see, e.g.,][for comprehensive
descriptions]{Townsend2003b,Mathis2013LNP}, even if they rotate up to a large
fraction of their critical rate \citep{Ouazzani2017}.
\begin{figure}
\begin{center}
\rotatebox{270}{\resizebox{3.3cm}{!}{\includegraphics{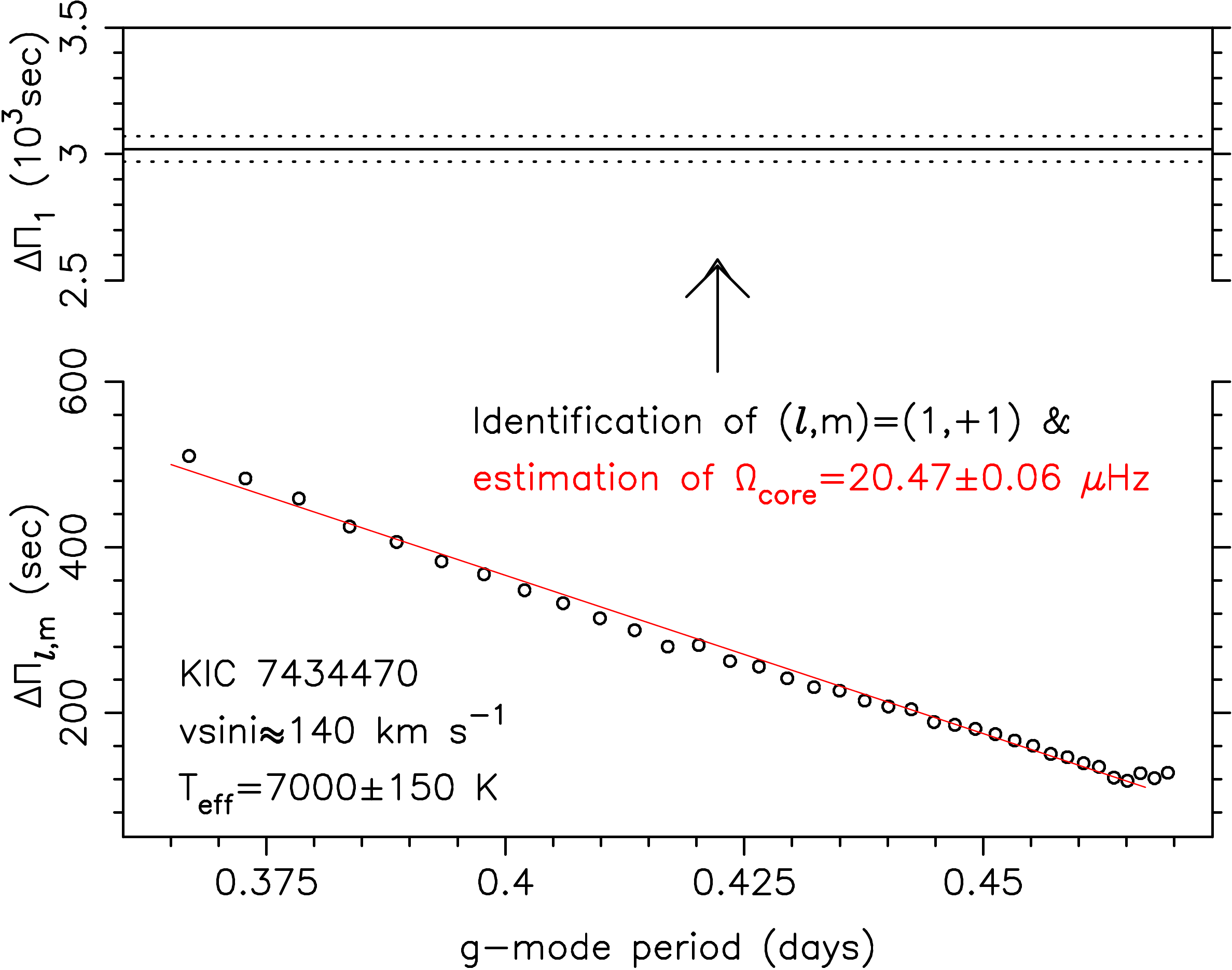}}}\hspace{0.3cm}
\rotatebox{270}{\resizebox{3.3cm}{!}{\includegraphics{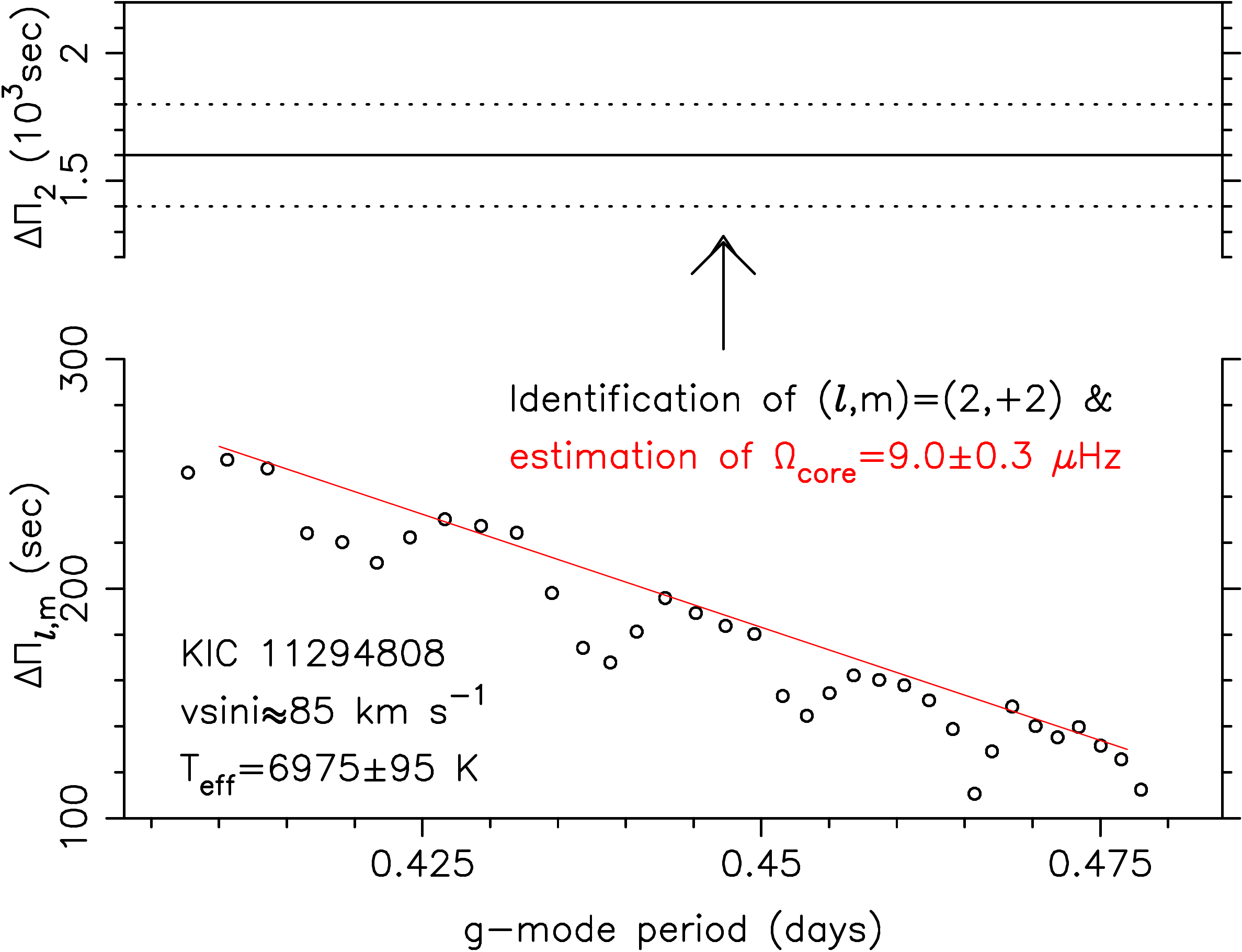}}}\\[1cm]
\caption{\label{fig-tilt} Period spacings versus periods for g~modes of
  consecutive radial order observed in two intermediate-mass F-type stars from
  4-year {\it Kepler\/} light curves (circles, errors smaller than the symbol
  size). These patterns deviate from a constant period spacing that would occur
  in the absence of rotation.  The slope of the observed pattern (red line) is
  caused by the Coriolis acceleration in the near-core region and allows to
  derive $\Omega_{\rm core}$ and to identify dominant $(l,m)$ through the application of
  Eqs\,(\ref{spacing-Tassoul}) and (\ref{spacing-TAR}), as indicated by the full
  lines in the upper panels (dotted lines indicate the uncertainty range for
  $\Delta\Pi_{l}$).  The occurrence of mode trapping allows to deduce the star's
  age and its chemical mixing properties, $D_{\rm mix}(r\!>\!r_{\rm cc})$, after
  estimation of $\Omega_{\rm core}$. Figures produced from data in
  \citet{VanReeth2015,VanReeth2016}.}
\end{center}
\end{figure}

In a non-rotating star with a convective core, the periods of low-degree,
high-order g~modes have an asymptotic spacing given by
\begin{equation}
\Delta \Pi_l \equiv \Pi_{l,n} - \Pi_{l,n-1}= {\Pi_0 \over \sqrt{l(l+1)}} \; ,
\label{spacing-Tassoul}
\end{equation}
where
\begin{equation}
\Pi_0 = 2 \pi^2 \left( \int_{r_1}^{r_2} N {\dd r \over r} \right)^{-1} \; ,
\end{equation}
with $r_1$ and $r_2$ the turning points of $N(r)$, indicative of the boundaries
of the convective region(s) that define the g-mode propagation cavity
\citep{Tassoul1980,Smeyers2010}. Typical values for $\Delta\Pi_1$ for B-type
g-mode pulsators range from 5000\,s to some 12000\,s for $M\in [3,8]\,$M$_\odot$
\citep{Degroote2010,Moravveji2015}. These values decrease as $l$ increases,
according to Eq.\,(\ref{spacing-Tassoul}).  \citet[][]{VanReeth2016}
computed $\Delta\Pi_l$ for dipole and quadrupole modes of F-type pulsators, varying
$(M,X,Z,X_c,\alpha_{\rm ov})$ and obtained $\Delta\Pi_1\in [2500,3500]$\,s
and $\Delta\Pi_2\in [1200,2200]$\,s for non-rotating stars.

Inclusion of the Coriolis force in the TAR, assuming uniform rotation, leads to
a similar period spacing pattern:
\begin{equation}
\Delta \Pi_{l,m} = {\Pi_0 \over \sqrt{\lambda_{l,m,\mathscr{S}}}} \; ,
\label{spacing-TAR}
\end{equation}
with $\lambda_{l,m,\mathscr{S}}$ the eigenvalue of the Laplace tidal equation
for the g mode with quantum numbers $(l,m)$ and $\mathscr{S}$ the spin parameter
$\mathscr{S}\equiv 2\,\Omega_{\rm core}/\omega_{nlm}$. Hence the observed period
spacing pattern of a series of g~modes with consecutive radial order $n$ and
dominant mode numbers $(l,m)$, allows one to simultaneously identify $(l,m)$ and
estimate the near-core rotation frequency $\Omega_{\rm core}$.  The procedure is
illustrated in Figure\,\ref{fig-tilt} for the gravito-inertial modes detected in
the {\it Kepler\/} single stars KIC\,7434470 and KIC\,11294808
\citep{VanReeth2016}.  From the observed diagnostics, it is found that
KIC\,7434470 is a $M\simeq 1.4\,$M$_\odot$ star in its early main sequence
evolution with $\Omega_{\rm core}/\Omega_{\rm crit}=62\%$, while KIC\,11294808
is a $M\simeq 1.9\,$M$_\odot$ star fairly close to the TAMS rotating nearly
uniformly \citep[$\Omega_{\rm core}/\Omega_{\rm env}=98\%$,][]{VanReeth2018} at
59\% of its critical rate (Mombarg et al., in preparation).  Their difference in mass
and evolutionary stage is not only revealed by the values of $\Delta\Pi_l$, but
as well by the difference in mode trapping properties: the $\mu$-gradient zone
of KIC\,11294808 is fairly extensive as its convective core has had a long time
to shrink, while KIC\,7434470 is in the mass range of having a growing
convective core early in its evolution and hence does not reveal mode
trapping. In this way, the period spacing patterns such as those in
Figure\,\ref{fig-tilt} make it possible to estimate $(M,X,Z,X_c)$ after
derivation of $\Omega_{\rm core}$ \citep{Aerts2018a}, as well as put constraints
on $D_{\rm mix}(r\!>\!r_{\rm cc})$ if mode trapping is observed as for KIC\,7434470
in Figure\,\ref{fig-tilt}.  The absence of mode trapping as in the case of
KIC\,11294808 shown in Figure\,\ref{fig-tilt}, only allows one to deduce a lower
limit of $D_{\rm mix}(r\!>\!r_{\rm cc})$.
Asteroseismic modeling of gravito-inertial modes has been put into practice
for single and binary B- and F-type g-mode pulsators on the basis of 4-year {\it
  Kepler\/} light curves by \citet{Moravveji2015,Moravveji2016},
\citet{VanReeth2016}, \citet{SchmidAerts2016}, \citet{Guo2017},
\citet{Papics2017}, \citet{Ouazzani2017}, \citet{Kallinger2017}, and
\citet{Szewczuk2018} at various levels of detail.  

\citet{Mathis2009} generalized the TAR to include general differential 
rotation both in radius and latitude. This allows to derive an improved 
expression compared to the one in
Eq.\,(\ref{spacing-TAR}) for the period spacing pattern. His formalism was
implemented and exploited by \citet{VanReeth2018} to study the sensitivity of
gravito-inertial modes to differential near-core rotation in a sample of 37
intermediate-mass main 
sequence stars included in Figure\,\ref{plot-rotation} discussed
below.  It was found that differential rotation can only be measured when period
spacing patterns for different degrees $l$ are detected or when the surface
rotation rate is measured from rotational modulation or p~modes. The stars for
which this is fulfilled all have surface-to-core rotation ratios between 0.95
and 1.05.  
\citet{Prat2017,Prat2018} go beyond the TAR by deriving an asymptotic period
spacing for axisymmetric sub-inertial gravito-inertial waves and a general
asymptotic theory for gravito-inertial waves propagating within a general
differential rotation both in $r$ and $\theta$, respectively. These theoretical
advances have yet to be applied to interpret data. Also 2D oscillation
codes have been built to compute oscillations of rapidly rotating stars using
either realistic 1D spherical stellar models or 2D deformed, often polytropic,
models \citep[e.g.][]{Dintrans2000,Reese2006,Ballot2010,Ouazzani2017}.

\subsubsection{Current Status of Asteroseismic Rotation Rates}

\begin{figure}
\begin{center}
\rotatebox{270}{\resizebox{5.5cm}{!}{\includegraphics{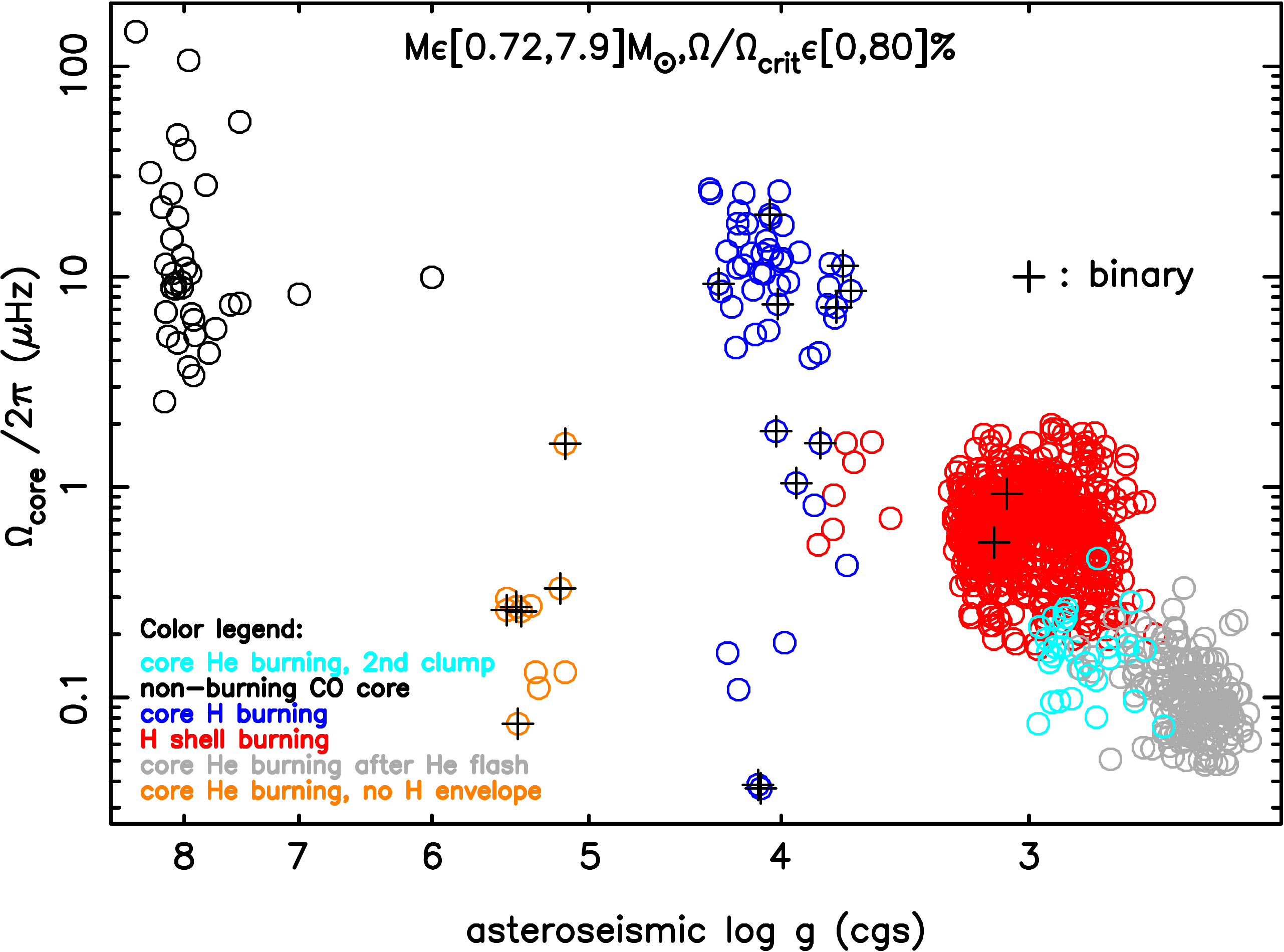}}}\\[3.cm]
\rotatebox{270}{\resizebox{5.5cm}{!}{\includegraphics{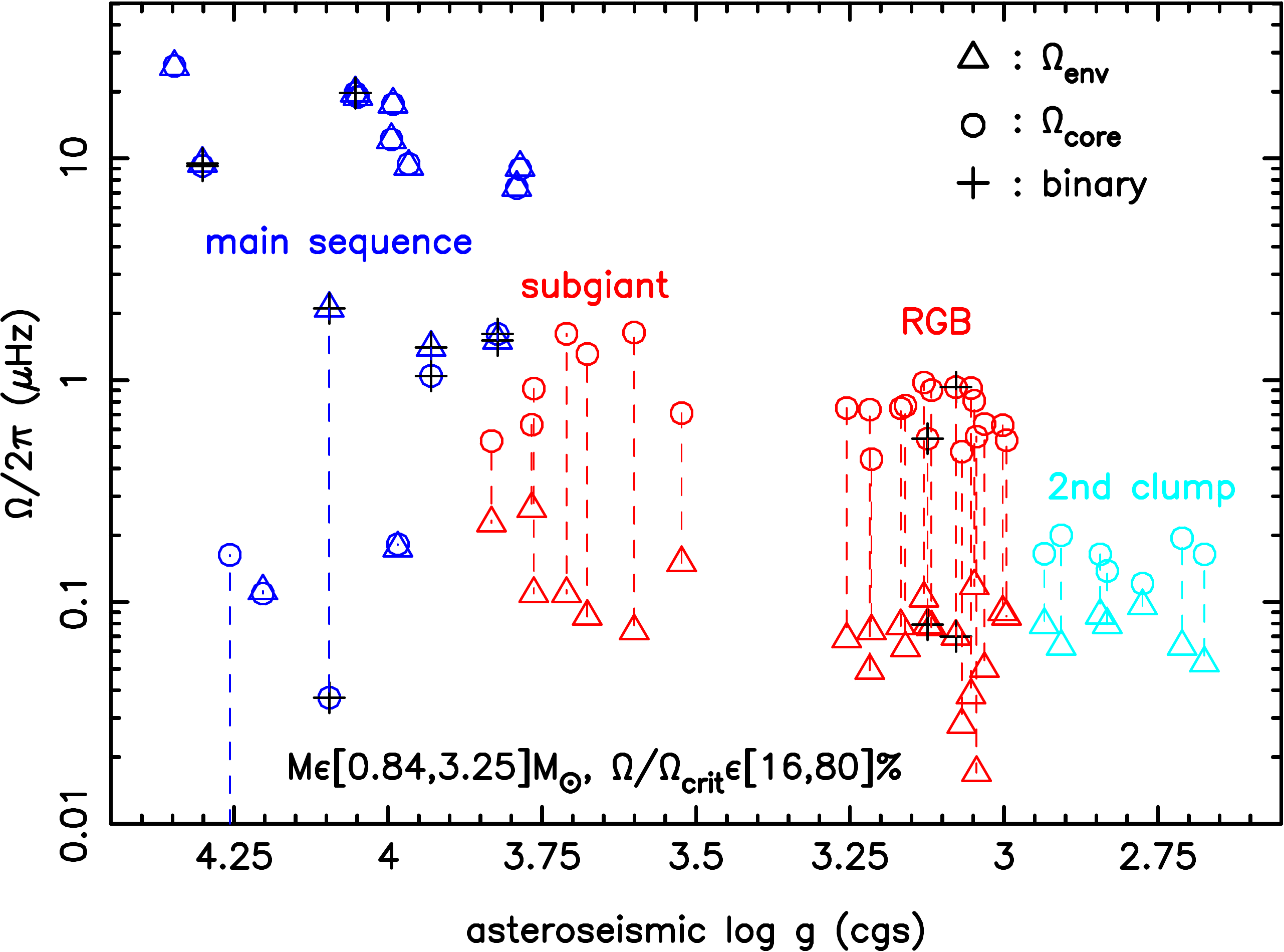}}}\\[1.5cm]
\caption{\label{plot-rotation} Top: 1210 stars with an asteroseismic estimate of
  $\Omega_{\rm core}$, $M$, and $R$ (dd.\,1/8/2018) for the indicated coverage of mass and
  critical Roche rotation rate as a function of their asteroseismic $\log\,g$,
  where the latter is a proxy of the evolutionary stage. Typical errors for
  $\Omega_{\rm core}$ are listed in Table\,\ref{precisions}, while those of
  $\log g$ range from 0.01 to 0.3\,dex and are omitted for optimal
  visibility. Bottom: 45 stars with an estimate of $\Omega_{\rm env}$ as
  well.  Figure produced from data in
  \citet{Mosser2012,Deheuvels2012,Deheuvels2014,Deheuvels2015,Kurtz2014,Saio2015,Triana2015,Triana2017,SchmidAerts2016,Murphy2016,Moravveji2016,DiMauro2016,Reed2016,Aerts2017,Sowicka2017,Hermes2017a,Guo2017,Kallinger2017,Gehan2018,Buysschaert2018,Szewczuk2018,Saio2018b,Beck2014,Beck2018,VanReeth2018}, 
Johnston et al.\ (submitted), and Mombarg et al.\ (in preparation).}
\end{center}
\end{figure}

The upper panel of Figure\,\ref{plot-rotation} shows the asteroseismic estimates
of $\Omega_{\rm core}$ and $\log g$ for 1210 stars available in the literature
(dd.\ 1 August 2018), based on space photometry that led to suitable identified
modes allowing to derive these two quantities.  The rotation rates were derived
as illustrated in Figures\,\ref{fig-splitting} and \ref{fig-tilt}, while the
asteroseismic masses and radii were deduced from scaling relations of damped
p~modes or from forward modeling of coherent g~modes as discussed in
Sections\,\ref{forward-low} and \ref{forward-high}.
These 1210 stars
cover the entire evolutionary sequence from core hydrogen burning to the white
dwarf remnant phase, for birth masses ranging from 0.72 to 7.9\,M$_\odot$ and
rotation rates from essentially zero up to 80\% of the critical Roche
frequency.
Several of the main sequence stars are in binaries; these have been
indicated as such. While the sample of main sequence stars is not
representative in mass, age, rotation rate and binarity, the red clump samples
leave no doubt that their helium burning cores have similar angular momentum
to subdwarfs and white dwarfs. This result is found for both red clump and
secondary clump stars, and is hence independent of having undergone a helium
flash or not.

Further, the upper panel of Figure\,\ref{plot-rotation} reveales that
$\Omega_{\rm core}$ of all red giants is about two orders of magnitude lower
than standard theory of angular momentum transport predicts
\citep[e.g.][]{Marques2013,Goupil2013,Cantiello2014,Ouazzani2018}.  
This means the cores
have to lose angular momentum to their overlying envelopes much more efficiently
than any current theory predicts. Irrespective of their birth mass or binarity,
all red giants reveal that their progenitors are subject to yet unknown physical
processes that are able to extract angular momentum from their core in an
efficient manner.  It has been suggested that the efficiency of the unknown
angular momentum transport process(es) is mass dependent \citep{Eggenberger2017}
but it remains unclear if a dichotomy occurs between stars born with and without
a convective core \citep{Tayar2013} or whether the angular momentum transport
increases gradually in efficiency with birth mass.
\begin{marginnote}[]
\entry{Red clump star} a red giant on the horizontal branch that initiated its
  core-helium burning in degenerate matter through a helium flash
  ($M \simkl 2.3\,$M$_\odot$)

\entry{Secondary red clump star} a red giant that initiated its core-helium burning
in a non-degenerate state ($M \simgr 2.3\,$M$_\odot$)
\end{marginnote}

In the lower panel of Figure\,\ref{plot-rotation}, we shown the 45 stars with a
measurement of both $\Omega_{\rm core}$ and $\Omega_{\rm env}$ from space
photometry.  These cover a narrower mass range of $[0.84,3.25]$\,M$_\odot$ but
an almost equally broad spread in $\Omega/\Omega_{\rm crit}$ as the upper panel.  For
some of the measurements in the lower panel, inversion methods similar to those
of helioseismology 
have been used to derive
$\Omega(r)$ throughout the star, instead of (kernel-)averaged values for the
core and envelope.  From those inversion 
studies, it has been found that the dependence of
the results for $\Omega_{\rm core}$ on the adopted 1D equilibrium models through
the use of the rotation kernels is modest and not important in the context of
angular momentum transport theory during stellar evolution (see, e.g.,
\citet{Deheuvels2012,Deheuvels2014,Deheuvels2015,DiMauro2016,Triana2017} for
sub- and red giants and \citet{Triana2015} for a B-type star). The lower panel of
Figure\,\ref{plot-rotation} reveals that intermediate-mass main sequence stars rotate
nearly uniformly (unless they are member of a binary) and that red clump
stars are efficient in reducing the level of differentiality of the rotation
acquired during the subgiant and red giant phases, when they have a radiative
core surrounded by a hydrogen burning shell.

Asteroseismic determination of rotation for main sequence low-mass
stars with damped p~modes  cannot reveal
$\Omega_{\rm core}$, but only $\Omega_{\rm env}$.  By combining their
rotationally split p~modes with a high-precision spectroscopic measurement of
$v\sin\,i$ and stellar models or with the frequency of rotational modulation
detected in the {\it Kepler\/} data, \citet{Benomar2015} and \citet{Nielsen2017}
achieved the envelope and surface rotation rates of 27 such single stars,
covering a mass range of 1.0 to 1.6\,M$_\odot$. This revealed nearly uniform
envelope rotation with values $\Omega_{\rm env} \in [0.4,3.2]\mu$Hz, in 
agreement with the spectroscopic $\Omega_{\rm surf}$ values.

Angular momentum transport by IGW triggered by a convective core in
intermediate-mass stars \citep{Rogers2013} and by the envelope in low-mass stars
\citep{Charbonnel2005}, offers a natural explanation of these observational
findings, but there may be other mechanisms as well. In any case, candidate
mechanisms must operate and be able to reduce the level of differentiality of
$\Omega(r)$ during both core hydrogen and core helium burning for stars with a
convective core \citep{Deheuvels2015,Aerts2017}.  Cores of low-mass stars spin
up during the subgiant and RGB phases, to reach a level up to ten
times the envelope rate, irrespective of whether they belong to a binary or not.

Figure\,\ref{plot-rotation} shows $\Omega_{\rm core}$ rather than the angular
momentum of the stellar core,
$J_{\rm core}\equiv M_{\rm core}\cdot\Omega_{\rm core}\cdot R^2_{\rm core}$,
with $M_{\rm core}$ and $R_{\rm core}$ the mass and radius of the stellar core,
respectively.  The reason is that the asteroseismically derived
$\Omega_{\rm core}$ is (quasi) model-independent, while the computation of
$J_{\rm core}$ requires knowledge of the core mass and core radius, which are
model-dependent. A discrepancy of two orders of magnitude between the observed
angular momentum of young neutron stars and white dwarfs on the one hand, and
the theoretical predictions for $J_{\rm core}$ of their progenitor stars on the
other hand, was previously reported \cite[see Figure\,1 in][]{Langer2012}.
Asteroseismology has now made it clear that this mismatch must be sought in the
early phases of stellar evolution, at least for low- and intermediate-mass
stars.  Their cores lose far more angular momentum than predicted by 
theory, so an efficient mechanism is required to transport it from the core to
the outer envelope. Despite the limited sample of main sequence stars in
Figure\,\ref{plot-rotation}, it is found that the angular momentum transport
happens efficiently when the star has a convective core.  High-precision
asteroseismic measurements of both $\Omega_{\rm env}$ and $\Omega_{\rm core}$
from space photometry are currently lacking for high-mass stars, and in
particular for blue supergiants. As discussed in the future prospects, this will
be remedied in the coming years.

\begin{summary}[SUMMARY POINTS FROM OBSERVATIONS]
\begin{enumerate}
\item Asteroseismology makes it possible to determine the interior
  rotation of stars, provided that they offer suitable nonradial
  oscillations to do so;
in the cases of rotational splitting detected for mixed or high-order g~modes, this is 
achieved in a model-independent way. 
\item {\it Kepler\/}/K2 space photometry of $\mu$mag precision and 4-year
  duration led to the core rotation rates of more than 
1200 stars covering the mass range $[0.7,8]$\,M$_\odot$,
rotation frequencies up to 80\% of the critical Roche frequency, and
evolutionary phases from core
hydrogen burning to the white dwarf remnant phase.
\item Low- and intermediate-mass main sequence stars 
with an asteroseismic measurement of the core
rotation rate, rotate nearly uniformly.
\item Stars lose a large fraction of their initial core angular momentum between the 
core hydrogen burning phase and the end of core helium burning.
\item The asteroseismically measured 
angular momentum of the core of single low- and intermediate-mass 
stars in the core helium burning phase agrees with
  the angular momentum of subdwarfs and white dwarfs.
\item Current theory of angular momentum transport falls short by one-to-two
  orders of magnitude to explain the asteroseismic core rotation rates of evolved low- and
  intermediate-mass stars.
\end{enumerate}
\end{summary}
\noindent Armed with these recent asteroseismic constraints on $\Omega(r)$, 
we now turn our
attention to new developments in the theory and simulation of angular momentum
transport, offering the reader a historical overview of the relevant facets of
this topic and stressing the need for improvements.

\section{ANGULAR MOMENTUM TRANSPORT}

In the standard treatment of stellar structure and evolution, radiative zones
are assumed to be motionless. However, as proven by helio- and asteroseismology,
they are the seed of dynamical processes, which act on secular time scales to
transport angular momentum and mix chemical elements. 
Here, we focus only on angular momentum transport.
Four main processes are
identified \citep[][]{Zahn2013,Mathis2013LNP}: meridional circulation,
turbulence driven by instabilities, magnetism, and internal waves. Together
these mechanisms give rise to angular momentum transport within
radiative interiors of stars. Note, however, that the term ``rotational
transport/mixing'' is sometimes used by the stellar evolution community as terminology
for these processes altogether.

\subsection{Hydrodynamical Meridional Circulation: 
from Eddington Sweet to Zahn's Model}

Historically, the large-scale meridional circulation, which stirs stellar
radiative zones, was ascribed to the deformation of the isobars by the
centrifugal acceleration, the Lorentz force if there is a magnetic field, and
the tidal force if there is a close companion. In that case, the radiative flux
is no longer divergence-free \citep{vonZeipel1924} and should be balanced by
heat advection. In the treatment derived by
\cite{Eddington1925} and \cite{Sweet1950}, the meridional circulation velocity was
thus linked to the ratio of the centrifugal acceleration (and the other
perturbing forces) to the gravity. The characteristic time scale of this flow
was derived by \citet{Sweet1950} and named the Eddington-Sweet time scale:
$t_{\rm ES}=t_{\rm KH} \displaystyle{\frac{GM}{\Omega^2R^3}}$, where
$t_{\rm KH}=\displaystyle{\frac{GM^2}{RL}}$ is the Kelvin-Helmholtz time scale.
These results predicted that rapid rotators should be well mixed because of this
circulation and should modify their evolution to the giant branch compared to
non-rotating stars; this is not observed. This was then improved by \cite{Mestel1953}, who accounted
for the mean molecular weight (i.e., the $\mu$-gradients), which reduces these
circulation effects.

However, the fact that meridional circulation advects angular momentum was
ignored. After a first transient phase lasting an Eddington-Sweet time, the star
settles into an asymptotic regime. At this stage, the circulation is driven by
structural adjustments, torques apply to the star and, when we restrict to the
simplest case ignoring fossil magnetic fields and internal waves, internal
stresses such as those related to shear-induced turbulence occur
\citep{Zahn1992,Rieutord2006}. On one hand, if the star loses angular momentum
through a wind, the circulation adjusts to transport angular momentum towards
the surface \citep{Zahn1992,MaederZahn1998,MathisZahn2004,Decressin2009}. The
induced rotation resulting from the advection of the angular momentum is then
non-uniform and a baroclinic state sets in, where the temperature varies with
latitude along the isobar. On the other hand, if the star does not exchange
angular momentum with its environment, the advection by the circulation balances
the internal stresses (this is the so-called gyroscopic pumping). In the case of a uniform rotation without any turbulent
transport, magnetism, and waves, the circulation thus dies out
\citep{Busse1982}. In the case of late-type stars it is hence the loss or
gain of angular momentum that drives the circulation, rather than the amplitude
of the angular velocity or the related centrifugal acceleration. 

The loop of angular momentum transport in stellar radiative zones by large
scale circulation can be identified as follows
\citep[see][]{Rieutord2006,Decressin2009}:
first, meridional currents are sustained by the torques applied at the stellar
surface, by internal stresses such as the viscous ones related to turbulence,
and by structural adjustments; next, the temperature relaxes to balance the
advection of entropy by the meridional circulation; finally, because of the
baroclinic torque induced by the latitudinal distribution of temperature
fluctuations on the isobar, a new differential rotation profile is established
through the so-called thermal-wind balance. This shear may again generate
turbulence, closing the loop.

\subsection{Hydrodynamical Instabilities and 
Convective Penetration/Overshoot}

Differential rotation in convectively stable radiative zones can induce a large
diversity of hydro- and magneto-hydrodynamical instabilities, as discussed
below. The turbulence induced by these instabilities can then transport angular
momentum.  It was already stressed above that mixing at convective-radiative
interfaces is critical for both angular momentum and chemical evolution.

The hydrodynamical instabilities, which can take place in the bulk of stellar
radiative zones, can be classified in three main families: the Rayleigh-Taylor
instability, the baroclinic instabilities, and the shear instabilities. The
Rayleigh-Taylor and baroclinic instabilities draw their energy from the
potential energy built up by the entropy stratification and the centrifugal
acceleration, while the shear instabilities' energy originates from the kinetic
energy of the medium.

\subsubsection{The Rayleigh-Taylor Instability}
The first major instability for a differentially rotating fluid is the so-called
Rayleigh-Taylor instability, which is directly linked to the Coriolis
acceleration. It was first studied by Lord \citet{Rayleigh1916} in the case of
a homogeneous and inviscid flow and by \cite{Taylor1923} in the case of a
viscous fluid. Let $s$ be the distance to the axis of rotation in
cylindrical coordinates. The so-called Rayleigh or epicyclic frequency
$N_{\Omega}$ is defined as
\begin{equation}
N_{\Omega}^{2}=\frac{1}{s^3}\frac{{\rm d}}{{\rm d}s}\left[\left(s^2\Omega\right)^2\right].
\end{equation}
The differentially rotating fluid is stable if $N_{\Omega}^{2}>0$ and unstable
if $N_{\Omega}^{2}<0$. The Rayleigh frequency corresponds to the response time to the
restoring Coriolis acceleration and therefore plays an analogous role to the
Brunt-Va\"is\"al\"a frequency ($N$) when considering fluctuations in 
the entropy 
and chemical
stratification. To connect these frequencies, 
the Brunt-Va\"is\"al\"a frequency can conveniently be written as a sum of
two frequencies corresponding to the restoring of entropy ($N_{T}$) and chemical
stratification ($N_{\mu}$), respectively:
\begin{equation}
N^{2}=N_T^2+N_{\mu}^2=\frac{g}{H_P}\left[\delta
\left(\nabla_{\rm ad}-\nabla\right)+\varphi\nabla_\mu\right],
\end{equation}
where we have introduced $\nabla=\displaystyle{\frac{\partial \ln T}{\partial \ln
    P}}$, $\nabla_{\rm ad}=\displaystyle{\left(\frac{\partial \ln T}{\partial
      \ln P}\right)_S}$, $\nabla_{\mu}=\displaystyle{\frac{\partial \ln
    \mu}{\partial \ln P}}$, $\delta=\displaystyle{\left(\frac{\partial \ln
      \rho}{\partial \ln T}\right)_{P,\mu}}$, and
$\varphi=\displaystyle{\left(\frac{\partial \ln \rho}{\partial \ln
      \mu}\right)_{P,T}}$, where $\rho$ is the density, $P$ the pressure, $T$
the temperature, $S$ the entropy, and $\mu$ the mean molecular weight. Following
\citet{Solberg1936} and \citet{Hoiland1941}, the Rayleigh criterion can be generalized to
take into account the stratification. A differentially rotating stratified
region is stable if $N^2+N_{\Omega}^{2}>0$ and if 
the specific angular momentum per unit mass, $j=s\Omega$, grows from the pole
towards the equator on each isentropic surface.

\subsubsection{Baroclinic Instabilities}
In the case of a general differential rotation law rather than 
uniform or cylindrical rotation $\left(\Omega\left(s\right)\right)$, stars are
in a baroclinic situation, where the entropy gradient and the gravity are not
aligned \citep[e.g.][]{Zahn1992}. 
This configuration can lead to unstable axisymmetric
displacements. The growth of these displacements is hindered by the
stable stratification, but this stabilizing effect gets reduced on
 small scales by thermal diffusivity.  Therefore, these
instabilities need to be described in the presence of thermal
diffusivity $K_T$ and viscosity $\nu$, keeping in mind that $K_T$ is several orders of
magnitude larger than $\nu$. 
In a stellar radiative region with an unstable differential rotation
and a stable entropy stratification, the effect of heat and momentum diffusion
is to weaken the effect of stratification by the ratio $\nu/K_T$. Such a situation
where heat is diffused faster than momentum leads to a so-called double
diffusive instability like semi-convection if $N_T^2<0$ and $N_\mu^2>0$ and
the thermohaline instability if $N_T^2>0$ and
$N_\mu^2<0$. \cite{GoldreichSchubert1967} and \cite{Fricke1968} identified this
instability, now known as the Goldreich-Schubert-Fricke (GSF) instability, 
and derived the
instability criteria:
\begin{equation}
\frac{\nu}{K_T}N_T^2+N_\Omega^2<0\quad\hbox{or}\quad
\vert s \partial_z\Omega^2\vert>\frac{\nu}{K_T}N_T^2,
\end{equation}
where $z$ is the coordinate along the rotation axis. The first criterion is the
Solberg-Hoiland condition modified by the presence of dissipative processes,
while the second one is directly linked to the baroclinicity of the star when
$\partial_z\Omega\ne0$. These criteria can be generalized in the case where the
chemical stratification is taken into account. The first instability criterion then
becomes
\begin{equation}
\frac{\nu}{K_T}N_T^2+\frac{\nu}{K_\mu}N_\mu^2+N_\Omega^2<0,
\end{equation}
where $K_\mu$ is the molecular diffusivity, which is generally of the same 
order of magnitude as the viscosity. In the case of a strong stabilizing
chemical stratification, the GSF instability can be inhibited. However, 
another axisymmetric instability, the Axisymmetric-BaroClinic-Diffusive (ABCD)
instability is triggered because of the action of heat diffusion
\citep{KnoblochSpruit1983}. The instability criterion then becomes
\begin{equation}
\frac{\nu}{K_T}\left(N_T^2+N_\mu^2\right)+N_\Omega^2<0.
\end{equation}
Finally, non-axisymmetric baroclinic instability can be triggered if
\begin{equation}
\left|\frac{\partial\ln\Omega}{\partial\ln
  r}\right|>C\left(\frac{N}{\Omega}\right)^2\frac{H_\Omega\ {\rm
    min}\left(H_\rho,H_N,H_\Omega\right)}{r^2}, 
\end{equation}
where $H_X=\vert{\rm d}\ln X/{\rm d}r\vert^{-1}$, with
$X=\left\{\rho,N,\Omega\right\}$ while $C$ is a coefficient close to unity
\citep{Spruit1983,Zahn1983}. This development is favored in weakly stratified
regions such as layers close to convective/radiative interfaces.

All the instabilities can play a role in the transport of angular momentum
and the mixing of chemicals in stellar interiors \citep{HirschiMaeder2010}.  Usually they are implemented as a diffusion coefficient in 1D stellar evolution codes, but
their treatment and implementation is challenging
since their non-linear saturation is not properly understood nor modeled.

\subsubsection{Shear Instability} 
The second family of hydrodynamical instabilities in stellar radiative zones
concerns those triggered by vertical and horizontal shears \citep{Zahn1992}. In
a non-stratified case, the necessary condition to have an instability is that
the velocity profile $V$ has an inflexion point, i.e.,
${\rm d}^2 V/{\rm d}x^2=0$, where $x$ is the direction along which the
instability is considered. In the case of a stably stratified differentially
rotating region, the situation becomes more complex
\citep{Zahn1992,Mathis2018}. Indeed, the competition between the destabilizing
action of the shear and the stabilizing buoyancy and Coriolis forces has to be
examined in detail.  

If we first consider the effect of a vertical
shear (i.e., along the entropy stratification), assuming as a first step that
$2\Omega\!<\!\!<\!N$, the instability is obtained in an adiabatic configuration
if $R_{i}=N^2/{\widehat{\cal S}}^2<1/4$, where we have introduced the Richardson number ($R_{i}$)
and the shear ${\widehat{\cal S}}={\rm d}V/{\rm d}z$. This situation is not reached generally in
stellar interiors. However, as in the case of the GSF instability, the action of
heat and momentum diffusion should be taken into account. More specifically,
since heat is diffused more rapidly than momentum, the heat diffusion weakens
the stabilizing action of the entropy stratification. The instability criterion
then becomes \citep[e.g.][]{Zahn1992}
\begin{equation}
\frac{N^2}{{\widehat{\cal S}}^2}\frac{v\ell}{K_T}<R_{i,c},
\end{equation}
where $R_{i,c}$ is the critical Richardson number, while $v$ and $\ell$ are
characteristic turbulent velocities and length scales, respectively. This leads
to the related turbulent transport coefficient along the vertical direction
\begin{equation}
D_{V,V}=\frac{R_{i,c}}{3}K\left(\frac{r\sin\theta}{N}\frac{{\rm d}\Omega}{{\rm d}r}\right)^2,
\end{equation}
which has been validated by recent high-resolution direct numerical simulations
in a Cartesian geometry \citep[e.g.][]{Prat2013}. From now on $D_{i,j}$
designates the eddy diffusivity corresponding to the turbulent transport along
the direction $i$ because of the instability of the shear in the
direction $j$. The turbulent motions sustained by the instability of the vertical
shear are three-dimensional. As a consequence, they also trigger turbulent
transport along the horizontal direction orthogonal to the entropy
stratification \citep{Mathis2018}. In stably stratified rotating fluids, the
turbulent transport is anisotropic. Indeed, the buoyancy inhibits turbulent
motions in the vertical direction while the Coriolis acceleration acts as the
restoring force along the horizontal direction. In stellar interiors, where
$2\Omega\!<\!\!<\!N$, this leads to a horizontal turbulent transport coefficient
derived by \cite{Mathis2018}:
\begin{equation}
D_{H,V}=\frac{N^4\tau^2}{2\Omega^2}D_{V,V}\quad\hbox{with generally}
\quad D_{H,V}\!>\!\!>\!D_{V,V},
\end{equation}
where $\tau$ is a characteristic turbulent time scale that depends on the vertical
shear and on rotation. We refer to \cite{Mathis2018} for a detailed
discussion of this time scale.

Finally, a horizontal shear (i.e., $\Omega\left(\theta\right)$) can trigger
finite-amplitude instabilities \citep{Zahn1983,Richard1999,Dubrulle2005}. This
was the first source of turbulence invoked to lead to efficient horizontal
transport of momentum and forms the corner stone of the shellular rotation
approximation derived by \cite{Zahn1992}.  Few and incomplete prescriptions have
been derived as of today for the related horizontal turbulent transport
coefficient $D_{H,H}$. The first one, based on dimensional arguments was
proposed by \cite{Zahn1992} but it was found by \cite{Maeder2003} that it leads
to configurations along stellar evolution where the condition
$D_{V,V}\!<\!\!<\!D_{H,H}$ is not respected. Subsequently, \cite{Maeder2003} and
\cite{Mathis2004} derived two other prescriptions. The first one was based on
energetic considerations on the horizontal shear while the second one was using
results obtained on the turbulent transport in a non-stratified Couette-Taylor
experiment \citep{Richard1999,Dubrulle2005}. Both prescriptions lead to larger
values for $D_{H,H}$ with $D_{V,V}\!<\!\!<\!D_{H,H}$ but new theoretical,
numerical, and experimental efforts should be pursued to provide a robust
ab-initio prediction.

\subsubsection{\label{penetration}Convective Penetration and Overshoot} 
A major source of turbulence that should be considered when studying stellar
evolution is convective penetration/overshoot at the border(s) of
convective/radiative zones. Let us follow \citet{Zahn1991}, who provided a
coherent physical picture of these processes. First, turbulent stellar
convection zones are the seed of large-scale coherent structures commonly
referred to as plumes. Because of their inertia, they penetrate into adjacent
stellar radiative zones. The key physical control parameter is the Peclet number
defined as $P_e=v_c\,l_c/K_T$, where $v_c$ and $l_c$ are the characteristic
convective velocity and length scale, respectively.  Two regions are
identified. In the first one, $P_e>1$, i.e., the flow dynamics is driven by the
advection and the plume keeps its identity. The plumes penetrate into the stable
region and they render it nearly adiabatic over a penetration distance
$d_{\rm pen}$ while they are decelerated by the buoyancy force. When the Peclet
number drops below unity, $P_e<1$, the thermal diffusion operates faster than
advection and the temperature gradient adjusts from adiabatic to radiative in an
overshoot region with thickness $d_{\rm over}$.

To guide stellar modelers in their quest to describe
convective penetration/overshoot, \cite{Vialletetal2015} proposed three
regimes. First, for plumes with $P_e\le1$, which are only able to mix
composition without affecting the entropy structure, an exponential diffusion
coefficient as proposed by \cite{Freytagetal1996} can be adopted. For plumes
with $Pe\ge1$, the entropy and the chemicals are both efficiently mixed and one
can assume that $\nabla=\nabla_{\rm ad}$ while the chemical composition is
assumed to be mixed instantaneously. Finally for plumes with $P_e\!.>\!\!>\!1$,
turbulent entrainment of mass by convective flows \citep{Fernando1991} occurs.  
Another way to improve the quantification of convective penetration and
overshoot is by numerical simulations, where strong efforts have been undertaken
to follow the whole coverage in Peclet number
\citep[e.g.][]{Browning2004,Meakin2007,Viallet2013,Rogers2013,Brun2017}.

Nowadays, the overshoot region can be probed observationally by investigating
the effect of the different prescriptions on observed and identified oscillation frequencies
\citep[e.g.][]{Deheuvels2016,Constantino2017,Pedersen2018}. This is a promising way to
constrain the physical properties and free parameters of the prescriptions at
the boundaries of convective and radiative layers.  Yet another challenging task
which asteroseismology may be able to take up in the future is to determine the
dependence of convective penetration/overshoot on rotation and magnetic fields.

Secular 1D stellar modeling has shown that the combined action of hydrodynamical
turbulent transport and of the advection by the meridional circulation can
explain some properties of massive stars \citep[e.g.][]{MeynetMaeder2000}, such
as surface abundances \citep[e.g.,][]{Hunter2008}.  However, as explained in
Section\,\ref{classical}, this is only the case for a fraction of stars. Moreover,
since these phenomena are unable to explain the observed rotation profile of the
solar radiative interior \citep[e.g.][]{Pinsonneault1989,Turck2010,Mathis2018}
nor the level of differential rotation of low- and intermediate-mass stars
recently found by asteroseismology shown in Figure\,\ref{plot-rotation}.  Therefore, other
physical mechanisms must be sought.  Magnetic fields and IGW are obvious
candidates.


\subsection{Angular Momentum Transport by Magnetism}

As already highlighted in Section\,\ref{classical}, the magnetic fields of
low-mass stars are highly variable dynamo fields, while those observed in about
10\% of the intermediate- and high-mass stars are strong stable structured
large-scale fossil fields. A dynamo generated field in the convective
core of main sequence stars would not penetrate the extended overlying radiative zone
\citep{MacGregor2003}. The typical decay time of a fossil magnetic field is
given by $\tau_D\approx R^2/\eta$, where $\eta$ is the magnetic
diffusivity. This is approximately $10^{9}$ to $10^{10}$ years, so a fossil
magnetic field would be present throughout the stars' evolution, if acted on
solely by non-turbulent resistivity.

While the nature and cause of magnetism in stars varies, we are concerned with
the angular momentum transport mediated by it, rather than with its
origin so we do not discuss the generation mechanism.  Moreover, since
angular momentum transport within convection zones is efficient due to the
turbulent motions (magnetic or otherwise), we focus on angular momentum transport
by magnetic fields within radiative regions.

Probably the simplest form of angular momentum transport by magnetic
fields comes from the work of \citet{Ferraro1937}.  Let us start with the
magnetic induction equation:
\begin{equation} 
\dxdy{\bf{B}}{t}=\nabla\times\left(\bf{u}\times\bf{B}\right)+\eta\nabla^2\bf{B},
\end{equation}
assuming that the magnetic diffusivity $\eta$ is constant. 
We consider a rotating star with rotation rate $\Omega$ and with a magnetic
field $\bf{B}$ that 
is symmetric about the rotation axis.  We decompose the magnetic
field into poloidal and toroidal components, such that
${\bf B}={\bf B}_T+{\bf B}_P$, where $B_P\cdot\hat{\phi}=0$, $B_T=B_{\phi}\hat{\phi}$.
The only velocity component is that in the
azimuthal direction, such that ${\bf u}=r\Omega \hat{\phi}$, with $\hat{\phi}$ 
the unit longitudinal vector.  Finally, we
neglect magnetic diffusion. The azimuthal component of the magnetic
field then evolves according to: 
\begin{equation}
\dxdy{B_{\phi}}{t}=r\left(B_p\cdot\nabla\right)\Omega.
\label{MagRot}
\end{equation}
Assuming a steady state, this reduces to $\left(B_p\cdot\nabla\right)\Omega=0$,
which means that $\Omega$ must be constant on poloidal
field lines; this is known as Ferraro's iso-rotation law.  This homogenization
of angular velocity along field lines is mediated by Alfv\'en waves with a speed
$v_{A}=B/\sqrt{\mu_0\rho}$, where B is the magnetic field amplitude and $\mu_0$
the permeability of vacuum.  Different field lines could rotate at different
rates.  However, if differential rotation occurred, it would produce a magnetic
pressure between field lines, which would exchange angular momentum through
phase mixing and eventually these variations in rotation would decay, leaving
uniform rotation in the presence of a large scale poloidal field. In the case
where the fossil magnetic field connects to adjacent convective zones, some
differential rotation can be transmitted to the radiative zone along the field
lines \citep[e.g.][]{Strugarek2011}. In the case of strong
non-axisymmetric fields, rotation is uniform \citep{Spruit1999}.

If we do not consider a steady state, we can immediately see from
Eq.\,(\ref{MagRot}) that an initially dipolar field will be wound into a
toroidal field with its amplitude increasing in time. The presence of such a
toroidal field will reduce the efficiency of the angular momentum transfer
because the Alfv\'en waves that mediate angular momentum transfer are diverted
azimuthally.  Nevertheless, \cite{Mestel1987} found that even a weak magnetic
field could impose uniform rotation throughout the star, when considering the
solar radiative interior as an example.  A similarly weak field would be
necessary to impose uniform rotation within the radiative zone of a 
main sequence star.  Of course, the toroidal magnetic field cannot grow
indefinitely and is limited by magnetic instabilities.

\begin{marginnote}[]
\entry{MRI}{Magneto-Rotational Instability}
\end{marginnote}

One expects that the interiors of stars have some degree of radial differential
rotation because of the structural adjustments along their evolution and of
torques applied at their surface.  As discussed in Section\,\ref{1Dmodels}, this
is due to the continuous spin down of the stellar surface due to a magnetized
wind in the case of low-mass stars and a radiation-driven wind for high-mass
stars, while Figure\,\ref{plot-rotation} shows that low but significant levels of
differentiality are observed for intermediate-mass stars.  
In such a case, the
instabilities that the toroidal components of a magnetic field are likely subject to
are the Magneto-Rotational Instability
\citep[MRI,][]{Chandrasekhar1960,balbus1992} and/or the Tayler
\citep{tayler1975}/Pitts-Tayler \citep{pitts1985} instability.  These
instabilities were reviewed by \citet{Spruit1999} under stellar interior
conditions.

The MRI occurs in the presence of a weak field when the angular velocity of the
system decreases away from the rotation axis, while the angular momentum
increases. This instability derives its energy from the differential rotation
and has been studied extensively in the astrophysical disks community, where it
leads to turbulence within the disk and the requisite angular momentum transfer
to allow mass accretion onto the protostar.  It has received relatively little
attention within the stellar community because the instability is thwarted by
the stable stratification. The stability condition in the absence of diffusion
is:
\begin{equation}
q=-\frac{\partial \ln \Omega}{\partial \ln r} > \frac{N^2}{2\Omega^2},
\end{equation}
where $q$ is the differential rotation parameter.  For a typical main sequence star
the right hand side is $\sim\!10^5$, requiring an
unrealistically strong  radial differential rotation for
instability to occur.  However, at small scales, the stratification
can be offset by thermal diffusion, such that the condition becomes: 
\begin{equation}
q=- \frac{\partial \ln \Omega}{\partial \ln r} > \frac{N^2}{2\Omega^2}
\frac{\eta}{K_T}.
\end{equation}
Given that $\eta/K_T<\!\!< 1$ in stellar interiors, it
is not unreasonable to suppose this instability may proceed in some
regions of stellar radiative zones at some ages. \citet{Arlt2003} and \citet{Jouve2015} studied this instability in spherical shells and both
found the instability to occur with consequent increased angular
momentum transport.  However, both of those studies were carried out
for an unstratified gas, hence it is unclear how much those results
apply to stellar radiative interiors.  The MRI has also been studied
within the Sun, where it has been suggested to occur within the
tachocline \citep{Kagan2014}. 

\citet{Spruit1999} argued that within stellar interiors, the most likely instability
is the Tayler instability.  This kink-instability derives its energy from
the magnetic field itself and is most likely to occur near the poles.
The most unstable mode is the $m=1$ mode and the stability condition is:
\begin{equation}
\dxdy{\ln\left(B_T^2\sin\theta\cos\theta\right)}{\theta} > 0,
\end{equation}
which can often be satisfied in the polar regions.  The growth rate of the
instability depends on rotation and is proportional to
$\omega_A^2/\Omega$, where $\omega_A$ is the Alfv\'en frequency based on
the poloidal field strength.  Similar to the MRI this instability is
limited by stable stratification, which can be offset by thermal
diffusion at small scales.  Accounting for thermal and chemical
stratification, the instability criterion considered by \citet{Spruit1999} is:
\begin{equation}
\frac{\omega_A}{\Omega} >
\left(\frac{N^2}{\Omega^2}\frac{\eta}{\kappa}+\frac{N_{\mu}^2}{\Omega^2}\right)^{1/4}\left(\frac{\eta}{r^2\Omega}\right)^{1/4},
\end{equation} 
but a more complete criterion and detailed derivation of it can be found in
\citet{Zahn2007a}.  Building on the robustness of this instability,
\citet{Spruit2002} claimed that it could sustain a dynamo in the
stably stratified regions of stars.  In his description the toroidal field could
be generated by differential rotation acting on the initial poloidal field (the
standard ``$\Omega$'' effect; see Equation \ref{MagRot}). Once this toroidal field is sufficiently strong,
it becomes unstable to the Tayler instability.  The instability then re-produces
the latitudinal and radial field components required to close the dynamo loop.

Angular momentum transport by magnetic fields must ultimately be derived from
the Lorentz (Laplace) force in the azimuthal component of the momentum equation
\citep{Charbonneau1993}:
\begin{equation}
\rho r^2 \sin^2\theta \frac{{\partial}\Omega}{{\partial}t} =
\frac{1}{\mu_0}{\bf B}_{P}\cdot{\bf\nabla}
\left(r\sin\theta B_\phi\right)+
{\bf\nabla}\cdot
\left(\rho \nu r^2\sin^2\theta {\bf\nabla}\Omega\right),
\end{equation}
where only magnetic and viscous stresses are taken into account. If one
considers variations in radius to be larger than those in latitude, the Maxwell
stresses (or magnetic torques) that arise are $\sim\!B_rB_{\phi}/r$.
\citet{Spruit2002} assumes that the growth of the
field saturates when the generation of the field due to the instability
is balanced by the diffusion of the field by turbulent magnetic diffusion.  
The instability field strength is found by considering the largest
scale at which the instability can occur despite the restoring force
due to buoyancy and the smallest
scale at which the instability still grows in the presence of magnetic
diffusion.  Accounting for the reduction in the buoyancy force at small
scales due to thermal diffusion, this gives an estimate for the
instability field strength at saturation.  Coupling this to an
estimate for the relationship between azimuthal and radial components
of the field based on the length scale of the instability, gives an
estimate of the Maxwell stress $B_rB_{\phi}$, which contributes to angular momentum
transport assuming these fields are maximally correlated.  \citet{Spruit2002} goes on
to assume that this stress can be written as 
\begin{equation}
{\mathcal S}_{\rm m} \approx \frac{B_r B_{\phi}}{4\pi}=\rho \nu_{e} r \dxdy{\Omega}{r},
\label{magvisc}
\end{equation}
where $\nu_e$ is the effective viscosity associated with magnetism.  Given the
saturation field strengths for $B_r$ and $B_{\phi}$, $\nu_e$ can be
estimated from the local conditions of the rotation, differential rotation,
stratification (both thermal and compositional), and thermal diffusivity.  This
prescription was implemented into 1D stellar evolution codes
\citep{Maeder2003,Maeder2004,Heger2005}. It was found that this efficient
angular momentum transport mechanism could slow down the cores of massive stars
sufficiently to be more consistent with observations of pulsar rotation rates.
In addition, \citet{Eggenberger2005} demonstrated it can lead to the uniform
rotation observed in the solar radiative core.  This was taken as evidence that
the so-called Tayler-Spruit dynamo exists and transports angular momentum
according to the prescription described above and laid out in
\citet{Spruit2002}. However, \citet{Cantiello2014} demonstrated that it cannot
explain the core rotation of red giants shown in Figure\,\ref{plot-rotation}.

The arguments laid out by \citet{Spruit2002} were conceptually appealing and
could explain some lingering observational questions to do with angular momentum
transport.  However, there were several physical deficiencies associated with
the ansatz, as detailed in \citet{Zahn2007a}.  The most severe of these is that
it remains unclear if a Tayler-Spruit dynamo can occur within stellar radiative
zones and that, if it did, whether it would saturate as \citet{Spruit2002}
envisioned it.  Indeed, numerical results by \citet{Zahn2007a} indicate that,
while a Tayler instability does occur, it does not produce a dynamo. While small
scale radial and latitudinal fields are generated by the instability of the
toroidal field, this does not regenerate the large scale axisymmetric poloidal
field, which continuously decays in the simulations by \citet{Zahn2007a}.
Moreover, rather than the instability being saturated by diffusion, it is
saturated by the Lorentz (Laplace) force that reacts back on the differential
rotation.  Therefore, the amplitudes of the fields estimated by
\citet{Spruit2002} should be revised.  Finally, rather than angular momentum
transport behaving as an enhanced viscosity as in Eq.\,(\ref{magvisc}), the
disturbances to the magnetic field are propagated as Alfv\'en waves, which do
not enhance viscosity.  The numerical results by \citet{Zahn2007a} were in
conflict with those produced by \citet{Braithwaite2006}. Although the latter
author claimed to have found a dynamo, his simulations did not include explicit
magnetic diffusion and ran less than a magnetic diffusion time, which is
insufficiently long.  In addition, \citet{Braithwaite2006} permanently sustained
differential rotation in his simulations through a body force that does not
allow physical back-reaction of the Lorentz force, which generally acts against
shear.  Recent analytical and numerical work by \citet{Goldstein2018} showed
that the occurrence of the Tayler instability may depend on the physical
approximation used. We conclude that more work on the Tayler instability in
stellar radiative interiors is needed.
Despite these limitations of the Spruit ansatz, it remains the de-facto magnetic
prescription used in many 1D stellar evolution codes and is used
widely throughout the stellar evolution community.

\subsection{Angular Momentum Transport by Internal Gravity 
Waves}

\subsubsection{Wave Generation}
Angular momentum transport by IGW depends crucially on the properties of the
waves that are generated. Indeed, as discussed in the next section, their
propagation and dissipation depend sensitively on their frequency and length
scales.  Within stars propagative IGW can be generated by any disturbance to the
stably-stratified region.  The most common and efficient mechanisms of
generation are an adjacent turbulent convection zone and tidal forcing by a
companion star (or planet).

The tidal generation of waves was discussed in detail in \citet{Zahn1975}.  The
companion causes a disturbance at the convective-radiative interface that
generates IGW, which propagate away from the convection zone.
\citet{Goldreich1989} considered the case of intermediate- and high-mass main
sequence stars in which the energy of IGW propagates outward. However, the
ansatz is equally valid for stars with convective envelopes, in which case the
energy of the IGW propagates inward. The waves generated are of large scale
($l=2,m=2$ if coplanar, $m=1$ if inclined) and have a frequency equal to the
forcing frequency.  The dissipation of these waves through critical layers,
nonlinear breaking or radiative diffusion (see the next section) can lead to
efficient angular momentum exchange between the companion's orbit and the
star.

For single stars the predominant source of propagative IGW is their generation
by the convective core, although, as we pointed out in
Section\,\ref{Astero-age}, they can also be generated by a convective envelope
or by thin convection zones due to local opacity enhancements in radiative
envelopes (the $\kappa$-mechanism).  The generation
mechanism of IGW has been split into two sources: bulk excitation through
Reynolds stresses \citep[e.g.][]{Goldreich1990} and direct excitation through
plumes \citep[e.g.][]{Townsend1966,Schatzman1993}.  The formulation by
\cite{Goldreich1990}, which was originally developed for the excitation of
p~modes, has been the most widely adopted version
\citep{Kumar1997,Kumar1999,Lecoanet2013,Shiode2013}, despite the fact that it
neglects direct excitation of IGW by plumes, a likely significant source.  In
that formulation the generation of IGW is treated as an inhomogeneous wave
equation with a turbulent source term. \cite{Goldreich1990} argued that the
source term was dominated by the quadrupole term (Reynolds stresses) because the
monopole (non-adiabatic expansion/contraction of fluid) and dipole (buoyancy)
terms, while physically larger, virtually cancel each other.  While this
assumption and its consequences have been justified in the case of p~modes, it
is unclear whether it is valid in the case of IGW.  It has been demonstrated by
\citet{Lecoanet2015} that treating IGW generation as an inhomogeneous wave
equation with a source term does reproduce the results of self-consistent,
nonlinear numerical simulations of the generation of IGW in the context of
laboratory experiments. However, efforts should still be undertaken to formulate
the turbulent source term theoretically. 

In general, the spectra obtained by the theoretical studies based on the
\cite{Goldreich1990} formulation have a steep power-law dependence on
frequency.  In this formulation the convective turnover (or eddy) frequency is
generated with large amplitude, while higher frequencies are generated with
significantly lower amplitudes, with the energy in the waves scaling as
$E\propto \omega^{-a}$, with $a$ between 3 and 6.  This is in stark contrast
with numerical results by \citet{Rogers2010}, \citet{Rogers2013},
\citet{Alvan2014}, and \citet{Edelmann2017},
who all find power spectra that are much flatter, with frequency dependence
$E\propto \omega^{-1}$.  This difference has a significant impact on the
efficiency with which IGW can transport angular momentum and mix species, where
the theoretical results from \citet{Goldreich1990} lead to less efficient
transport and mixing than the numerical results.  While it is somewhat unclear
why there is such a difference between simulations and theory, there are a few
obvious contenders.  It is unclear whether the approximations made in
\citet{Goldreich1990} for p~modes are appropriate when applied to IGW, e.g., the
source term lacks plume excitation, the turbulent spectra are assumed
to be of Kolmogorov type, the effects of rotation and magnetism on
turbulence are ignored as is the lack of intermittency of convection.   In particular, the theoretical work by \cite{Goldreich1990} relies heavily
on the identification of a dominant turnover frequency within the convection
zone, but such a frequency does not occur in numerical simulations.  

In the theoretical plume model of excitation \citep[e.g.][]{Schatzman1993}, the waves
are generated following a Gaussian function in plume size and incursion time,
which results in a wave spectrum that is exponential in frequency and
wavelength.  This model has been revisited by \citet[][Figure\,3]{Pincon2017}, who
showed that this process results in a shallow energy generation at low frequency
and a steep one at high frequency.  Moreover, as the plume incursion time is
decreased, i.e., as the intrusion becomes more impulsive, the spectrum becomes
flatter in frequency, in line with numerical simulations.  It appears that theoretical plume models match numerical simulations better than those due to internal stresses \citep{Edelmann2018}, but clearly more work needs to be done.  

While the numerical simulations include many of the effects neglected in
theoretical work, they lack turbulence corresponding to stellar
regimes. Moreover, numerical constraints require enhanced damping of waves.
While there is no way to get to realistic levels of turbulence within the
convection zone, the simulations tuned to concrete stellar circumstances explain
the asteroseismic observations of main sequence stars shown in
Figure\,\ref{plot-rotation} \citep{Rogers2015}.  Some numerical simulations by
\citet{Rogers2013} force the convection harder to compensate for enhanced wave
damping. While this is not ideal there is no obvious better path forward if the
aim is to predict wave amplitudes at the stellar surface.  Before we move on to
wave propagation and dissipation, we point out that the combined effects of
tidally and convectively induced waves has not yet been considered, a lack that
should definitely be remedied in the future.

\subsubsection{Propagation and Dissipation}
In considering the propagation of IGW in stellar interiors, we first discuss
propagation of pure IGW, in the absence of rotation and magnetism.  In this
simplest case IGW propagate in stably stratified regions when $\omega < N$.
Whenever waves have high frequency $\omega$ close to $N$, the waves set up
standing modes due to internal reflection \citep{Alvan2015}.  However, here we
focus on low frequency travelling waves, which propagate throughout the interior
and are damped before internal reflection. In the limit $\omega\ll N$, the waves
have high radial order, which results in the vertical wavelength ($\lambda_v$)
being much smaller than the horizontal one ($\lambda_h$). In that case, the
horizontal velocities ($v_h$) are much larger than the vertical velocities
($v_v$), i.e. $\lambda_v/\lambda_h\sim\!\omega/N\sim\!v_v/v_h$.

There are three main ways to dissipate IGW: 1) linear thermal diffusion, 2)
occurrence of critical layers and 3) nonlinear breaking.  Thermal diffusion
depends sensitively on the wavelength and frequency of the waves, with the
damping length in the low-frequency regime
$l_d\propto K_T^{-1}k_h^{-3}\omega^4N^{-3}$.
Critical layers occur when the frequency of the wave is the same as the
local rotation frequency.  Formally, at such layers, the vertical wavelength
goes to zero.  However, nonlinear effects become important and the wave
effectively breaks, with a damping rate proportional to the square root of the
local Richardson number (see 3.2.3).  Finally, there are two
potential sources of nonlinear wave breaking: overturning of the stratification
(convective instability) and shear of the wave itself \citep[Kelvin-Helmholtz
instability][]{Thorpe2018}.  In stellar interiors, convective instability is
likely to happen only in regions where $N$ approaches 0, such as at the center
of low-mass stars and near convective-radiative interfaces.  Therefore, in the
bulk of stellar interiors, nonlinear wave breaking would take the form of a
Kelvin-Helmholtz instability \citep{Press1981}.  The simplest criterion for this
type of breaking is provided by
$\varepsilon\equiv v_h/\left(\lambda_h\omega\right)\sim\!1$. In intermediate- and
high-mass stars, waves generated at the convective core interface may reach
sufficient amplitude to satisfy this criterion.

\begin{figure}
\begin{center}
\rotatebox{0}{\resizebox{5.cm}{!}{\includegraphics{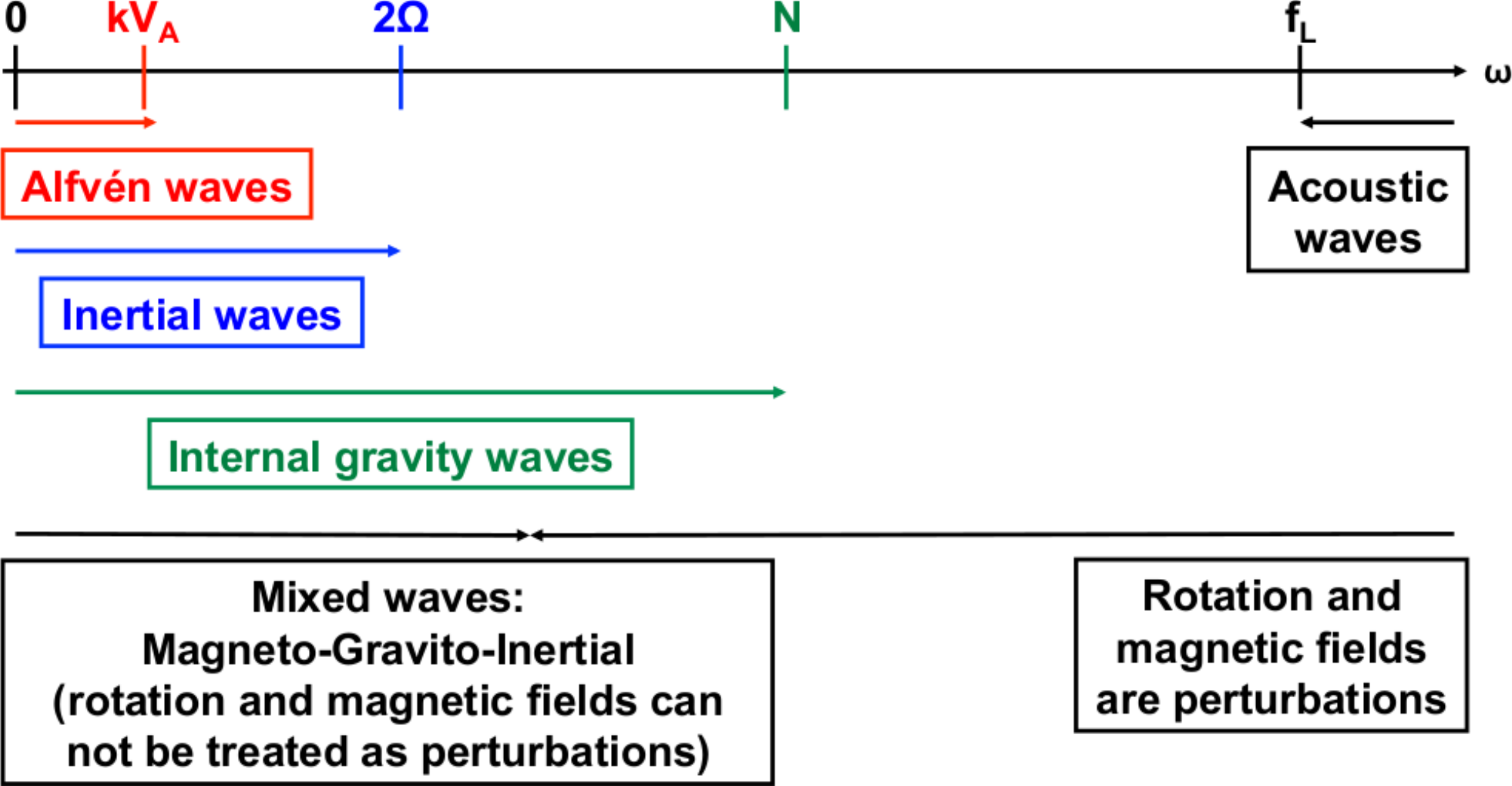}}}\vspace{0.5cm}
\caption{\label{waveplot} Wave types in a rotating and magnetized stably
  stratified radiative zone, with associated frequencies; $f_L$ is the Lamb
  frequency and $V_A$ the Alfv\'en speed, while $k$ is the wave vector.}
\end{center}
\end{figure}

In the case of rotating stars, waves are modified because of the Doppler effect
that creates a systematic frequency difference between prograde and retrograde
waves. This leads to thermal damping acting differently on prograde and retrograde
waves, thus sustaining a net transport of angular momentum.  In the case of
rapidly rotating stars, waves with frequencies close to $2\Omega$ are influenced
by the Coriolis acceleration. In the case of sub-inertial waves for which
$\omega<2\Omega$ (blue region in Figure\,\ref{waveplot}) the propagation
domain is restricted to the equatorial region. In the case of super-inertial
waves for which $\omega>2\Omega$ (green region in
Figure\,\ref{waveplot}) the waves continue to propagate in the full spherical
shell as in the non-rotating case since waves are less influenced by the
Coriolis acceleration \citep{LeeSaio1987,Dintrans2000}. In the case of
differential rotation, the behavior of waves is the same but with much more
complex propagation cavities than shown schematically in Figure\,\ref{waveplot}
\citep{Prat2018}.  This can have important consequences for the transmission of
energy from convective to radiative regions and on the transport of angular
momentum \citep{Mathis2009}.  In the sub-inertial regime, wave-induced
transport of angular momentum will be confined to the equatorial region.

Magnetic fields also affect the propagation and dissipation of IGW in stellar
interiors.  If the field is strong enough such that
$\omega_A=B/\left(r\sin\theta\sqrt{\mu_0\rho}\right)\sim\!\omega$, IGW can become
trapped \citep[e.g.,][]{Rogers2010,MathisdeBrye2011}.  In the case
of a toroidal field at the convective-radiative interface IGW propagation can be
completely blocked, while in the case of a poloidal field IGW are trapped along
poloidal field lines. Clearly, this affects low frequency waves more profoundly
than the high frequency standing modes used to diagnose stellar interiors.
These low frequency waves are the most relevant for the transport of angular
momentum.

\subsubsection{Transport}
In the absence of meridional flows and magnetic fields, the transport of angular
momentum by waves is given by a horizontal average of the azimuthal component of
the momentum equation \citep{Zahn1997}:
\begin{equation}
{\rho}\frac{{\rm d}}{{\rm
    d}t}\left(r^2{\overline\Omega}\right)=-\frac{1}{r^2}\partial_r\left(r^2\rho
  r \overline{\sin\theta v_r v_\phi}\right)+\frac{1}{r^2}\partial_r\left(\rho\nu
  r^4\partial_r{\overline\Omega}\right),
\end{equation}
where $v_r$ and $v_{\phi}$ are the radial and azimuthal velocities of the waves,
respectively, and overbars denote horizontal averages with
$\overline\Omega=\int_0^\pi\Omega\sin^3\theta{\rm
  d}\theta/\int_0^{\pi}\sin^3\theta{\rm d}\theta$ \citep{Zahn1992}.  This
equation shows that the mean zonal flow is accelerated/decelerated by the
divergence of the horizontally-averaged Reynolds stresses, often called the
Eliasen-Palm flux \citep{Eliassen1960} and damped by viscous dissipation.

In a global sense, angular momentum transport by IGW can couple convective and
radiative regions
\citep{Talon2002,Rogers2013,Tayar2013,Fuller2014,Rogers2015,Pincon2017}.  Hence,
there is major interest in these waves now that the asteroseismic results in
Figure\,\ref{plot-rotation} require efficient angular momentum coupling between
such regions.  The Doppler shift for IGW is
\begin{equation}
  \omega(r)=\omega_{\rm gen}-m[\Omega(r)-\Omega_{\rm gen}], 
\end{equation}
where $\omega_{\rm gen}$ and $\Omega_{\rm gen}$ are the frequency of excitation
and the angular velocity at the excitation radius, and $\omega(r)$ and
$\Omega(r)$ are the frequency and rotation rate in the local reference frame of
the wave \citep[][]{Fuller2014}. In the case of weak differential
rotation $\Delta\Omega\equiv\Omega(r)-\Omega_{\rm gen}<\omega_{\rm gen}/m$, IGW
sustain shear until $\Delta\Omega\ge\omega _{\rm gen}/m$. In that case of
stronger differential rotation, critical layers develop and these reduce the
differential rotation until $\Delta\Omega=\omega_{\rm gen}/m$.  From this we
immediately deduce that the largest differential rotation that can be tolerated
in the presence of IGW is $\Delta\Omega\sim\!\omega_{\rm gen}/m$.  For waves
generated by turbulent convection in a star, the wave spectrum of $\omega$ and
$m$ is unknown, which makes it difficult to estimate this limiting differential
rotation.

As an example, consider a main sequence star in which the convective core is
rotating faster than its radiative envelope, i.e., $\Delta\Omega<0$. Assuming
the convection generates prograde and retrograde waves equivalently, the
prograde waves ($m>$0) will be Doppler shifted to higher frequencies (if
$\Delta\Omega<0$), making them less susceptible to radiative damping as
explained in the previous section.  These can hence propagate further into the
radiative region, where they deposit positive angular momentum when they
dissipate.  On the other hand, retrograde waves ($m<$0) are shifted to lower
frequencies (if $\Delta\Omega<0$) and dissipate closer to the core, spinning
down the region, and, more importantly, transporting less angular momentum
because of the large density difference between the regions.  The outer regions
will hence spin up more than the inner regions will spin down.  In this way, the
convective core is coupled to the radiative envelope from the outside inward.
In that case, the transport enforces weak differential rotation.  In conclusion,
while waves can couple convective and radiative regions, it is expected that
these waves cause some positive or negative differential rotation within the
radiative regions of stars.  Therefore, as in the case of the advection by
meridional flows and of the magnetic stresses, the transport of angular momentum
by IGW cannot be modeled as a diffusive process.
 
\subsubsection{Application to stars}
The first studies of the transport of angular momentum induced by
stochastically-excited IGWs were carried out for the Sun
\citep{Schatzman1993,Kumar1997,Zahn1997,Talon2005}. \citet{Talon2005} showed
that IGW can couple the convective envelope and radiative interior as well as
bring about the uniform rotation of the radiative solar core. However, more
detailed analyses of IGW transport in the solar radiative region
\citep{Rogers2006,Denissenkov2008} indicated that these waves drive weak radial
differential rotation, which would be detectable by helioseismology, yet this
is not observed.  Therefore, while IGW can couple the convective and radiative
regions in the Sun, another mechanism seems necessary to explain the uniform
rotation profile of the bulk of the solar radiative interior up to
$0.25R_{\odot}$.  It is still unclear whether the core of the Sun rotates at the
same rate as the bulk of the radiative interior as both \citet{Garcia2007} and \citet{Fossat2017} show an
increased rotation rate in the core.

\begin{figure}
\begin{center}
\rotatebox{0}{\resizebox{6.5cm}{!}{\includegraphics{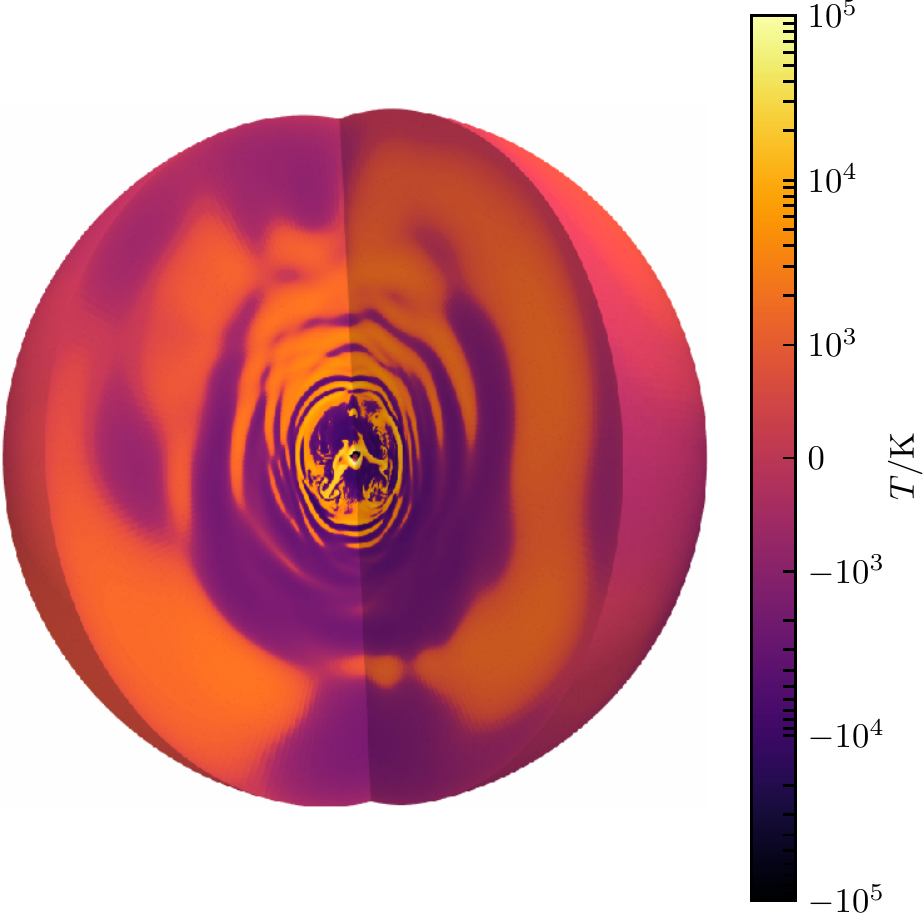}}}
\rotatebox{0}{\resizebox{6.2cm}{!}{\includegraphics{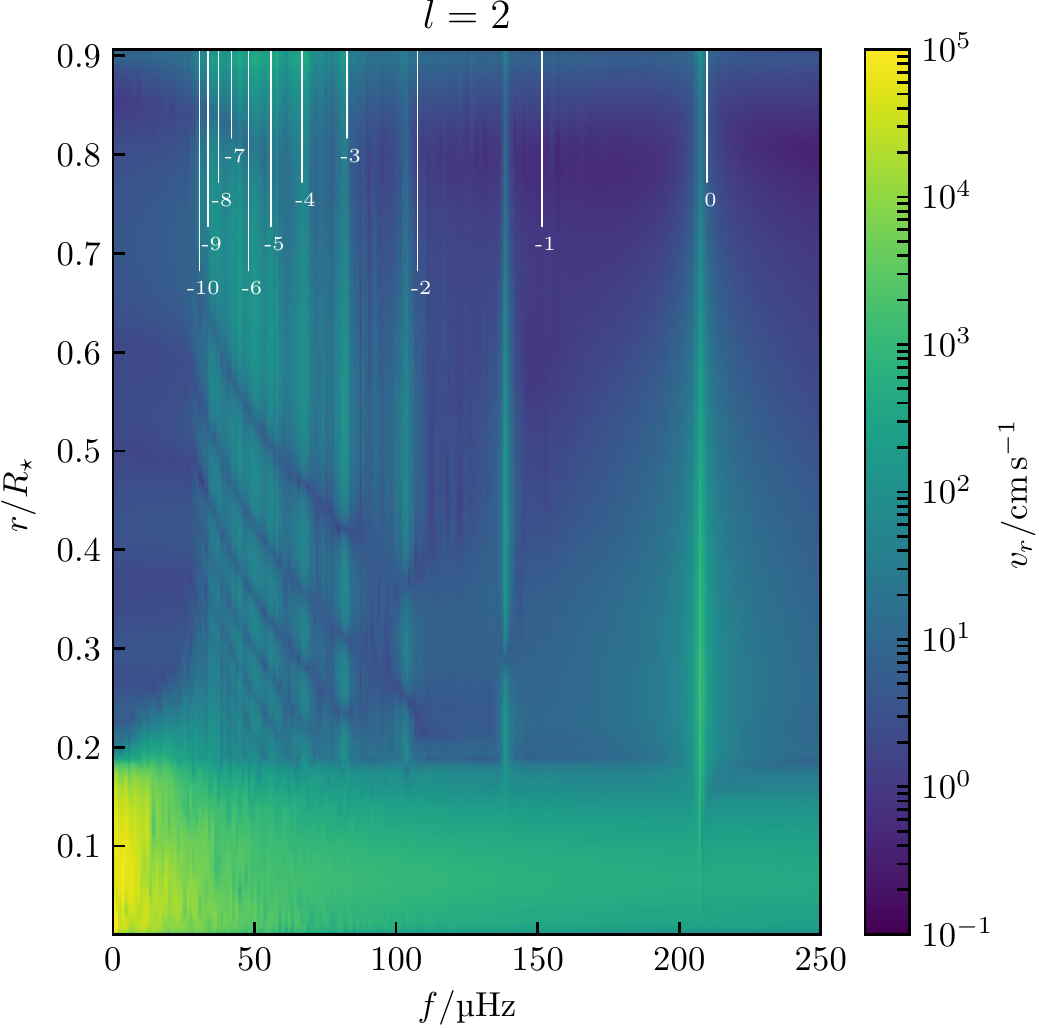}}}
\caption{\label{fig-3D} Left: 3D visualization of simulations of IGW from
  Edelmann et al.\ (submitted), where the color scale shows temperature
  fluctuations from the background state.  Right: Frequency spectra of the
  vertical velocity $v_r$ for IGW with $l=2$ throughout a 3\,M$_\odot$ ZAMS
  star. The vertical white lines are the frequencies of coherent $l=2$ modes of
  the stellar model for the indicated radial order.}
\end{center}
\end{figure}

In intermediate- and high-mass stars, IGW are likely to have more dynamical
consequences than in low-mass stars, because the waves propagate into a region
of decreasing density, allowing the amplitudes of some waves to be amplified.
Due to these increased amplitudes, some waves may break and interact with
critical layers leading to enhanced angular momentum transport.  Moreover, the
thermal diffusivity increases rapidly toward the surface of the star, which also
leads to efficient damping and angular momentum transport.  These effects have
been demonstrated in 2D simulations of an equatorial slice for a 3\,M$_{\odot}$
main sequence star \citep{Rogers2013,Rogers2015}.  Depending on the strength of
the convective flux used for the wave generation and of the level of rotation,
several regimes for the resulting differential rotation can be obtained. In
cases with high rotation and high forcing, the surface can spin faster than the
core in the same direction. In cases with high rotation and low forcing or low
rotation and low forcing, uniform rotation is obtained. In cases with low
rotation and high forcing, retrograde differential rotation can be
driven. Although incomplete in many ways, these numerical results are in
agreement with observations of differential rotation in the main sequence stars
shown in Figure\,\ref{plot-rotation}. This was a major motivation to generalize
the 2D simulations by \citet{Rogers2013} into 3D, with very similar results in
terms of wave spectrum properties and angular momentum transport. A snapshot of
the 3D temperature fluctuations is shown in the left panel of
Figure\,\ref{fig-3D}, while the comparison between the quadrupole component of
the generated IGW spectrum and the coherent quadrupole modes is shown on the
right (Edelmann et al., submitted). The tangential velocities and temperature
fluctuations resulting from these new large-scale long-term simulations are
shown in Supplemental Figure\,\ref{FigA1-A2}. Observed frequency values due to
coherent g~modes or IGW are dominantly determined by the near-core region for
intermediate-mass and high-mass stars, because their amplitudes and hence
probing power are dominant there. These fluctuations due to the IGW are
currently being used for the generation of synthetic light curves and
line-profile variations to evaluate their relevance in explaining low-frequency
power excess detected in space photometry \citep[and Bowman et al.\
submitted]{Tkachenko2014} and signatures of macroturbulence observed in
high-mass stars \citep[e.g.,][]{SimonDiazHerrero2014} for which oscillations
have been invoked as a physical explanation \citep{Aerts2009}.

In low-mass evolved stars, the efficiency of the transport of angular momentum
driven by IGW has been studied for the subgiant phase by \citet{Pincon2017}, who
showed that IGWs can extract enough angular momentum to reduce the differential
rotation induced by the stellar core contraction.  However, IGWs are not
efficient enough to explain the weak core rotation of red giants revealed in
Figure\,\ref{plot-rotation} because of the strong thermal dissipation induced by
the high stable stratification in the core \citep{Fuller2014,Cantiello2014}. As
an alternative \citet{Belkacem2015b,Belkacem2015a} showed that the angular
momentum driven by mixed modes can be a good alternative to explain the observed
low rotation rates.

\begin{summary}[SUMMARY POINTS FROM THEORY AND SIMULATIONS]
\begin{enumerate}
\item The zoo of hydrodynamical instabilities occurring in rotating stars is worth revisiting with the aim to calibrate them from asteroseismology; this will lead to a better understanding of their nonlinear properties and behavior as 
stars evolve. 
\item The various theoretical prescriptions of convective penetration and core
  overshooting can be evaluated from their signature on g~modes of single and
  binary intermediate-mass and high-mass stars; this will lead to an
  asteroseismically calibrated temperature gradient and mixing profile in the
  overshoot zone.
\item The theory of angular momentum transport by magnetism based on proper
  treatment of the Lorentz force should be revisited and included in stellar
  evolution codes. This can lead to predictions of observables to be tested
  against asteroseismology of magnetic pulsators of various kind
  \citep[e.g.,][]{Kurtz1990,Buysschaert2018}.
\item Simulations show that angular momentum transport by IGWs offer great
  potential to explain the asteroseismic results for $\Omega(r)$ observed in a
  variety of stars (Figure\,\ref{plot-rotation}). Strong efforts should be
  pursued to obtain good predictions for their excitation spectra, their
  propagation, and their damping in rotating (magnetized) stars.
\end{enumerate}
\end{summary}


\section{FUTURE OUTLOOK TO IMPROVE STELLAR MODELING}

Space asteroseismology opened a new window to the Universe, by providing direct
measurements of stellar interiors. As such, it revived the theory of stellar
structure and evolution, pointing out major challenges for theories that
remained uncalibrated so far but have been used in astrophysics for decades.  A
prominent shortcoming is a proper treatment of angular momentum transport in
stellar models. 

While the results in Figure\,\ref{plot-rotation} are certainly suggestive, a
drawback preventing us from deducing a causal connection between the core
hydrogen burning stars and their descendants is
the lack of an accurate age, aside from the far too small sample sizes.  The
asteroseismic gravity, while of good precision, is too rough a proxy for the
evolutionary status.  Therefore, we need to populate Figure\,\ref{plot-rotation}
with thousands of single and binary stars, covering the pre-main sequence up to
the supergiant phase, with suitable and identified nonradial oscillation modes.
Coming back to Figure\,\ref{spinner}, we have highlighted the major assets and
limitations of 1D stellar modeling, keeping in mind the observational
diagnostics and their level of precision in Table\,\ref{precisions} as well as
the current samples in Figure\,\ref{plot-rotation}.

Improvements in the theory, following a data-driven approach, are within reach
from existing {\it Kepler}/K2 and Gaia DR2 data.  The upcoming Gaia DR3 and TESS
data will reveal thousands of suitable pulsators for asteroseismology in the
Milky Way and in the Large Magellanic Cloud by the end of 2020.  However, probing of
$\Omega(r)$, $D_{\rm mix}(r)$ and other quantities in the deep stellar interior requires
gravito-inertial or mixed modes and their frequencies can only be deduced from
data with a duration of at least one year \citep{Aerts2018a}.  This can be
achieved for stars in the two Continuous Viewing Zones of TESS.  On a somewhat
longer term, the space data to be assembled with the ESA PLATO mission
\citep[][launch +2026]{Rauer2014} will deliver the necessary frequency precision
for samples of thousands of single and binary stars covering wide ranges in
mass, metalicity, and rotation, including pre-main sequence stars and stellar
clusters.  This big data revolution in stellar astrophysics will not only allow
us to populate Figure\,\ref{plot-rotation}, but also to replace the $\log\,g$ by a
high-precision seismic age estimate and to construct similar diagrams for other
quantities of critical importance for stellar interiors, such as the mixing
profiles, $D_{\rm mix}(r\!>\!r_{\rm cc})$, inside stars.  This revolution will also allow us to finally identify and better model the dominant sources of angular momentum transport.  

\begin{issues}[FUTURE ISSUES]
\begin{enumerate}
\item The way forward in stellar modeling is to apply an integrated data-driven
  approach connecting and developing synergies between observations, new
  theoretical developments, and new 3D\,(M)HD simulations according to the
  spinner in Figure\,\ref{spinner}. The {\it Kepler}/K2 and Gaia data sets
  largely remain to be exploited with this aim.
\item A theoretical formulation of the optimal turbulent source term for wave
  generation in a star should be sought.
\item 3D\,(M)HD simulations covering various nuclear burning phases should be
  performed; in order for them to be meaningful in terms of diagnostic
  predictions, these simulations must be global and large-scale in set-up,
  covering the whole star and have long-duration time bases of several months,
  so as to cover the observed low-frequency regime observed for stars.
\item An aspect of Figure\,\ref{spinner} that remained under-explored is
  tides in close binaries, the waves they trigger, and their impact on stellar
  interiors. The integration of tidal theory and binary asteroseismic modeling
  of observed stars should become a priority.
\item The development of 2D axisymmetric models taking into account the
  deformation from spherical symmetry due to the centrifugal force
  \citep[e.g.,][ESTER code]{Rieutord2016}, is of major importance to
  model stars rotating close to $\Omega_{\rm crit}$. Upgrading the current 2D
  models by including the chemical evolution of the star, envelope convection,
  and angular momentum transport is needed to exploit asteroseismic data of the
  fastest rotators.
\item Current samples of core hydrogen burning and hydrogen
  shell burning stars with an asteroseismic measurement of $\Omega(r)$ are not
  representative in terms of mass, rotation, and binarity; those samples need to be
  extended appreciably to become unbiased.
\item Combined $\mu$as astrometry, high-resolution high signal-to-noise
  spectroscopy, and asteroseismic measurements of $\Omega(r)$ and
  $D_{\rm mix}(r)$ throughout the star, for single stars, close binaries, and
  clusters constitutes an optimal route to achieve a better theory of stellar
  evolution.
\item Asteroseismic measurements of $\Omega(r)$ are hardly available for
  high-mass stars, and not at all for blue supergiants. The NASA TESS (launched
  April 2018) and ESA PLATO (to be launched 2026) missions offer the opportunity
  to assemble the required space photometry and to remedy this lack of stars for both
  Milky Way and Large Magellanic Cloud populations.
\end{enumerate}
\end{issues}


\section*{DISCLOSURE STATEMENT}
The authors are not aware of any affiliations, memberships, funding, or
financial holdings that might be perceived as affecting the objectivity of this
review.

\section*{ACKNOWLEDGMENTS}

The authors dedicate this review to Jean-Paul Zahn, for his education and
encouragements to all three of us; his interest and equal respect for all aspects of
our research -- observations, theory, and simulations -- was exceptional in the
astrophysics community and of major importance for our careers.

The authors proudly acknowledge the pleasant biannual group discussions and work
sessions among their postdoctoral researchers and PhD students the past few
years in Leuven, Newcastle, and Saclay, where the research in this
review was discussed.  They are also grateful for enlightening (email)
discussions with Nate Bastian, Paul Beck, Aaron Dotter, Jim Fuller, Thierry
Morel, Simon Murphy, Henk Spruit, Andrew Tkachenko, and Rich Townsend.  Clio
Gielen and Philipp Edelmann are thanked for having produced Figures 1, 6 and 7, and
Dominic Bowman, Cole Johnston, Charlotte Gehan, JJ Hermes, Joey Mombarg, Benoit
Mosser, May Gade Pedersen, and Timothy Van Reeth for having provided data
included in Table\,1 and Figures\,2 to 4 in electronic format.

The authors received funding from the European Research Council (ERC) under the
European Union's Horizon 2020 research and innovation programme (grant
agreements No.\ 670519: MAMSIE with PI Aerts and No.\ 647383: SPIRE with PI
Mathis), from Belspo PLATO grant at KU\,Leuven, from CNES PLATO grant at
CEA/DAp, and from the Science \& Technology Facilities Council (grant agreement
ST/L005549/1) as well as from NASA grant NNX17AB92G to PI Rogers.


\bibliographystyle{ar-style2.bst}
\bibliography{Astroph-Aerts-Mathis-Rogers}

\newpage


\section*{Online Material}

\begin{figure}[h!]
\begin{center}{}
\rotatebox{0}{\resizebox{6.3cm}{!}{\includegraphics{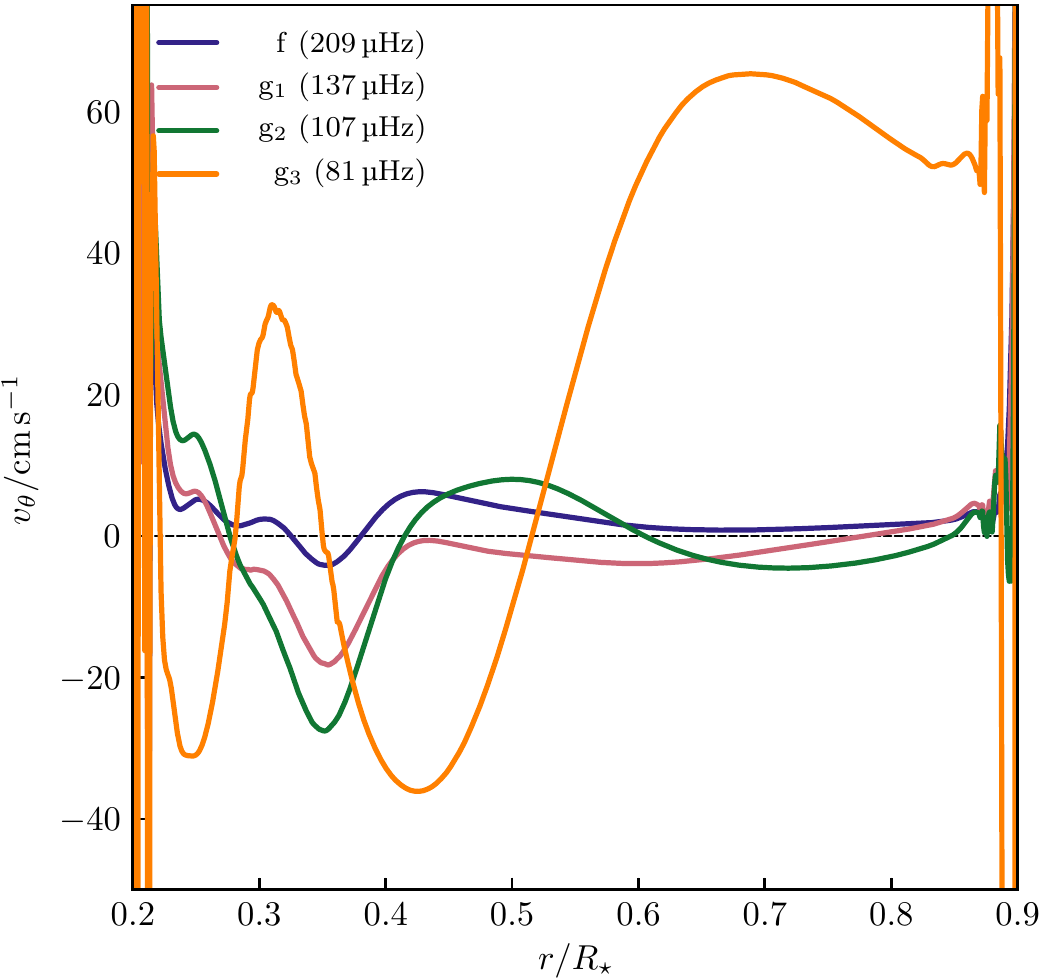}}}
\rotatebox{0}{\resizebox{6.3cm}{!}{\includegraphics{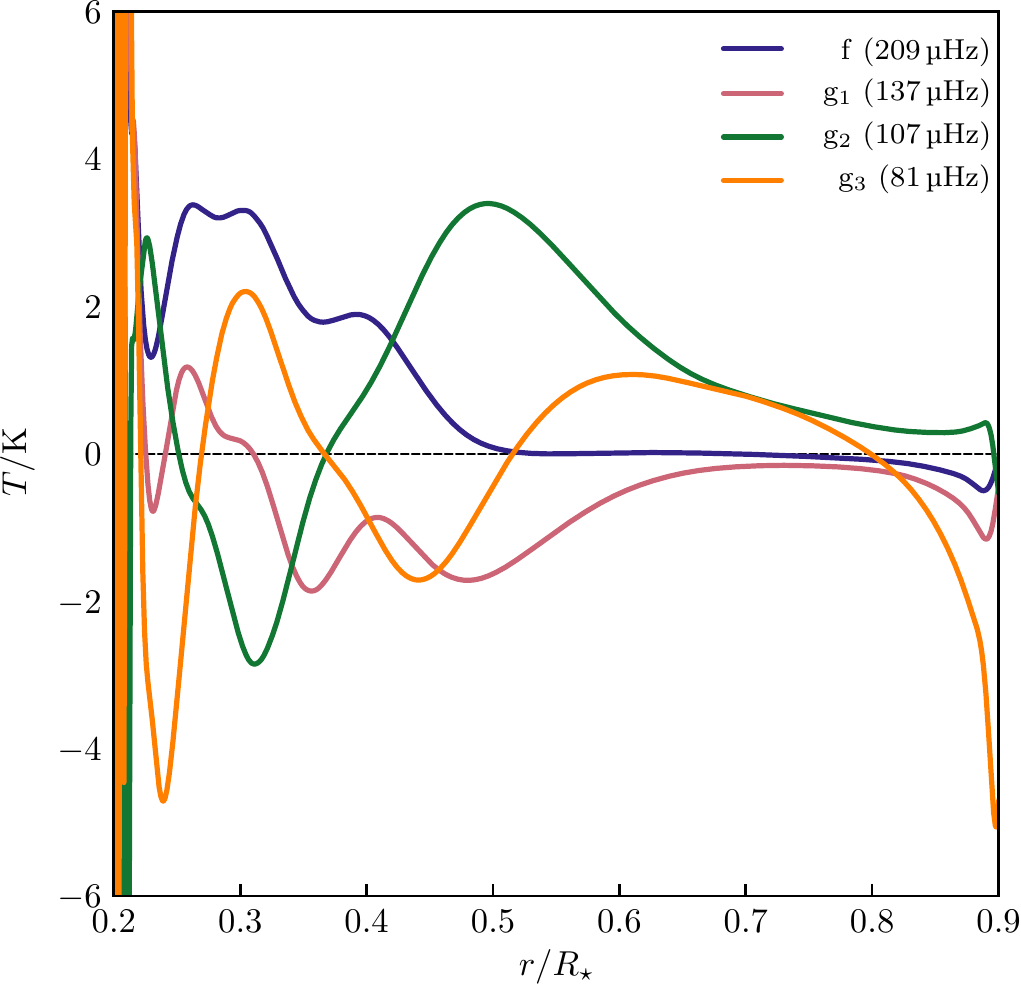}}}\\[1cm]
\caption{\label{FigA1-A2} The $\theta$ component of the horizontal velocity
  (left panel) and the temperature fluctuation (right panel) for four of the
  coherent quadrupole mode components present in the IGW spectrum generated by
  the core convection, with their indicated frequencies. The radial velocity
  component of these four modes is shown in Figure\,\ref{fig-3D} in the main
  text.  Figure produced from data in Edelmann et al.\ (submitted).}
\end{center}
\end{figure}

In Figure\,\ref{FigA1-A2}, we show $v_\theta$ and the temperature fluctuations
corresponding with a few of the quadrupole wave components of low order
occurring in the generated IGW spectrum for a 3\,M$_\odot$ main sequence star
computed by Edelmann et al.\ (submitted). From Figures\,\ref{fig-3D} and
\ref{FigA1-A2}, we deduce typical velocities of tens of cm\,s$^{-1}$ per
individual wave component, and accompanying temperature fluctuations of a few
K. These type of data are currently being used so simulate observables, such as
photometric light curves for the {\it Kepler\/} and TESS passbands and synthetic
high-resolution line-profile variations for comparison with  optical
spectroscopy of OB-type stars. In this way,  we will evaluate whether or not the
generated IGW spectra can explain the recent detections of low-frequency power
excess in such data of OB-type stars. Earlier, \citet{Aerts2009} already showed
that the collective effect of an ensemble of low-amplitude coherent g~modes of
degree $l=1,\ldots,10$ offers a physical explanation for the observed
line-profile broadening in some high-mass stars, generally called
macroturbulence \citep[e.g.][]{SimonDiazHerrero2014}. This is evidence that
g~modes are not only appealing as an explanation for the recent asteroseismic
measurements of $\Omega_{\rm core}$, but can also explain features detected in
time-series spectroscopy of high-mass stars that remain unexplained
otherwise. The ongoing simulation study on the basis of the IGW generated by
Edelmann et al.\ (submitted) offers a similar test for IGW instead of coherent
g~modes, from confrontation of the simulated diagnostics with space photometric
light curves as well as spectroscopic measurements of OB-type stars.\\[.5cm]

\end{document}